\documentclass[twocolumn]{aastex631}

\usepackage{}
\usepackage{gensymb}  
\usepackage{multirow}  

\shorttitle{SAPPHIRES: Extremely Metal-Poor Galaxies}
\shortauthors{Hsiao et al.}

\begin{document}

\title{SAPPHIRES: Extremely Metal-Poor Galaxy Candidates with $12+{\rm log(O/H)}<7.0$ at $z\sim5-7$ from Deep JWST/NIRCam Grism Observations}

\correspondingauthor{Tiger Hsiao}
\email{tiger.hsiao@cfa.harvard.edu}

\newcommand{\CfA}{\affiliation{Center for Astrophysics \text{\textbar} Harvard \& Smithsonian, 60 Garden Street, Cambridge, MA 02138, USA}}

\newcommand{\STScI}{\affiliation{Space Telescope Science Institute (STScI), 3700 San Martin Drive, Baltimore, MD 21218, USA}}

\newcommand{\JHU}{\affiliation{Center for Astrophysical Sciences, Department of Physics and Astronomy, The Johns Hopkins University, 3400 N Charles St. Baltimore, MD 21218, USA}}

\newcommand{\ESAAURA}{\affiliation{Association of Universities for Research in Astronomy (AURA), Inc.~for the European Space Agency (ESA)}}

\newcommand{\Austin}{\affiliation{Department of Astronomy, University of Texas at Austin, 2515 Speedway, Austin, Texas 78712, USA}}

\newcommand{\BGU}{\affiliation{Physics Department, Ben-Gurion University of the Negev, P.O. Box 653, Be'er-Sheva 84105, Israel}}

\newcommand{\kapteyn}{\affiliation{Kapteyn Astronomical Institute, University of Groningen, 9700 AV Groningen, The Netherlands}}

\newcommand{\CAB}{\affiliation{Centro de Astrobiología (CAB), CSIC-INTA, Ctra. de Ajalvir km 4, Torrejón de Ardoz, E-28850, Madrid, Spain}}

\newcommand{\Cambridge}{\affiliation{Kavli Institute for Cosmology, University of Cambridge, Madingley Road, Cambridge CB3 0HA, UK}}

\newcommand{\Cavendish}{\affiliation{Cavendish Laboratory, University of Cambridge, 19 JJ Thomson Avenue, Cambridge CB3 0HE, UK}}

\newcommand{\UCL}{\affiliation{Department of Physics and Astronomy, University College London, Gower Street, London WC1E 6BT, UK}}

\newcommand{\Arizona}{\affiliation{Department of Astronomy / Steward Observatory, University of Arizona, 933 N Cherry Ave, Tucson, AZ 85721}}

\newcommand{\TsingHua}{\affiliation{Department of Astronomy, Tsinghua University, Beijing 100084, China}}

\newcommand{\Chiba}{\affiliation{Center for Frontier Science, Chiba University, 1-33 Yayoi-cho, Inage-ku, Chiba 263-8522, Japan}}

\newcommand{\ICRR}{\affiliation{Institute for Cosmic Ray Research, The University of Tokyo, 5-1-5 Kashiwanoha, Kashiwa, Chiba 277-8582, Japan}}

\newcommand{\NAOJ}{\affiliation{National Astronomical Observatory of Japan, 2-21-1 Osawa, Mitaka, Tokyo 181-8588, Japan}}

\newcommand{\Sokendai}{\affiliation{Department of Astronomical Science, SOKENDAI (The Graduate University for Advanced Studies), Osawa 2-21-1, Mitaka, Tokyo, 181-8588, Japan}}

\newcommand{\IPMU}{\affiliation{Kavli Institute for the Physics and Mathematics of the Universe (WPI), University of Tokyo, Kashiwa, Chiba 277-8583, Japan}}

\newcommand{\Dawn}{\affiliation{Cosmic Dawn Center (DAWN), Denmark}}

\newcommand{\Cop}{\affiliation{Niels Bohr Institute, University of Copenhagen, Jagtvej 128, DK-2200 Copenhagen N, Denmark}}

\newcommand{\Tokyo}{\affiliation{Department of Astronomy, Graduate School of Science, the University of Tokyo, 7-3-1 Hongo, Bunkyo, Tokyo 113-0033, Japan}}

\newcommand{\Oxford}{\affiliation{Department of Physics, University of Oxford, Denys Wilkinson Building, Keble Road, Oxford OX1 3RH, UK}}

\newcommand{\CFC}{\affiliation{Cosmic Frontier Center, The University of Texas at Austin, Austin, TX 78712}}

\newcommand{\Westlake}{\affiliation{Department of Astronomy, Westlake University, Hangzhou 310030, Zhejiang Province, People’s Republic of China
}}


\author[0000-0003-4512-8705]{Tiger Yu-Yang Hsiao} \CfA \JHU \STScI \Austin \CFC

\author[0000-0002-4622-6617]{Fengwu Sun} \CfA \Westlake

\author[0000-0001-6052-4234]{Xiaojing Lin} \TsingHua \Arizona

\author[0000-0001-7410-7669]{Dan Coe} \STScI \ESAAURA \JHU

\author[0000-0003-1344-9475]{Eiichi Egami}
\Arizona

\author[0000-0002-2929-3121]{Daniel J.\ Eisenstein} \CfA 

\author[0000-0001-7440-8832]{Yoshinobu Fudamoto}
\Chiba \Arizona

\author[0000-0002-8651-9879]{Andrew J. Bunker} \Oxford

\author[0000-0003-3310-0131]{Xiaohui Fan}
\Arizona

\author[0000-0002-6047-430X]{Yuichi Harikane}
\ICRR

\author[0000-0003-4337-6211]{Jakob M. Helton}
\Arizona

\author[0000-0001-6874-1321]{Koki Kakiichi}
\Dawn \Cop

\author[0000-0003-4247-0169]{Yichen Liu} 
\Arizona

\author[0000-0003-3762-7344]{Weizhe Liu}
\Arizona

\author[0000-0002-4985-3819]{Roberto Maiolino} \Cambridge \Cavendish \UCL

\author[0000-0002-1049-6658]{Masami Ouchi}
\NAOJ \ICRR \Sokendai \IPMU

\author[0000-0003-3307-7525]{Wei Leong Tee}
\Arizona

\author[0000-0002-7633-431X]{Feige Wang}
\affiliation{Department of Astronomy, University of Michigan, 1085 S. University Ave., Ann Arbor, MI 48109, USA}

\author[0000-0003-0111-8249]{Yunjing Wu}
\TsingHua

\author[0000-0002-5768-8235]{Yi Xu}
\ICRR \Tokyo

\author[0000-0001-5287-4242]{Jinyi Yang}
\affiliation{Department of Astronomy, University of Michigan, 1085 S. University Ave., Ann Arbor, MI 48109, USA}

\author[0000-0003-3307-7525]{Yongda Zhu}
\Arizona





\newcommand{\LCDM}{$\Lambda$CDM}

\newcommand{\red}[1]{{\color{red} #1}}
\newcommand{\redss}[1]{{\color{red} ** #1}}
\newcommand{\redbf}[1]{{\color{red}\bf #1 \color{black}}}

\newcommand{\ny}{$\tilde {\rm n}$}
\newcommand{\about}{$\sim$}
\newcommand{\appr}{$\approx$}
\newcommand{\gt}{$>$}
\newcommand{\um}{$\mu$m}
\newcommand{\uJy}{$\mu$Jy}
\newcommand{\sig}{$\sigma$}
\newcommand{\Lya}{Lyman-$\alpha$}
\renewcommand{\th}{$^{\rm th}$}
\newcommand{\lam}{$\lambda$}

\newcommand{\tentothe}[1]{$10^{#1}$}
\newcommand{\tentotheminus}[1]{$10^{-#1}$}
\newcommand{\e}[1]{$\times 10^{#1}$}
\newcommand{\en}[1]{$\times 10^{-#1}$}
\newcommand{\cgsfluxunits}{erg$\,$s$^{-1}\,$cm$^{-2}$}
\newcommand{\linefluxunits}{\tentotheminus{20} \cgsfluxunits}

\newcommand{\logU}{$\log(U)$}
\newcommand{\logOH}{12+log(O/H)}

\newcommand{\sinv}{s$^{-1}$}
\newcommand{\kms}{km\,s$^{-1}$}

\newcommand{\footnoteurl}[1]{\footnote{\url{#1}}}

\newcommand{\tnm}[1]{\tablenotemark{#1}}
\newcommand{\super}[1]{$^{\rm #1}$}
\newcommand{\supa}{$^{\rm a}$}
\newcommand{\supb}{$^{\rm b}$}
\newcommand{\supc}{$^{\rm c}$}
\newcommand{\supd}{$^{\rm d}$}
\newcommand{\supe}{$^{\rm e}$}
\newcommand{\supf}{$^{\rm f}$}
\newcommand{\supg}{$^{\rm g}$}
\newcommand{\suph}{$^{\rm h}$}
\newcommand{\supi}{$^{\rm i}$}
\newcommand{\supj}{$^{\rm j}$}
\newcommand{\supk}{$^{\rm k}$}
\newcommand{\supl}{$^{\rm l}$}
\newcommand{\supm}{$^{\rm m}$}
\newcommand{\supn}{$^{\rm n}$}
\newcommand{\supo}{$^{\rm o}$}

\newcommand{\squared}{$^2$}
\newcommand{\cubed}{$^3$}

\newcommand{\sqarcmin}{arcmin\squared}

\newcommand{\supcomma}{$^{\rm ,}$}

\newcommand{\rhalf}{$r_{1/2}$}

\newcommand{\chisq}{$\chi^2$}

\newcommand{\Zgas}{$Z_{\rm gas}$}  
\newcommand{\Zstar}{$Z_*$}  

\newcommand{\per}{$^{-1}$}
\newcommand{\inv}{\per}
\newcommand{\Mstar}{$M^*$}
\newcommand{\Lstar}{$L^*$}
\newcommand{\phistar}{$\phi^*$}

\newcommand{\logM}{log($M_*$/\Msun)}

\newcommand{\LUV}{$L_{UV}$}
\newcommand{\MUV}{$M_{UV}$}

\newcommand{\Msun}{$M_\odot$}
\newcommand{\Lsun}{$L_\odot$}
\newcommand{\Zsun}{$Z_\odot$}

\newcommand{\Mvir}{$M_{vir}$}
\newcommand{\Mt}{$M_{200}$}
\newcommand{\Mf}{$M_{500}$}

\newcommand{\Ndotion}{$\dot{N}_{\rm ion}$}
\newcommand{\xiion}{$\xi_{\rm ion}$}
\newcommand{\logxiion}{log(\xiion)}
\newcommand{\fesc}{$f_{\rm esc}$}

\newcommand{\XHI}{$X_{\rm HI}$}
\newcommand{\XHII}{$X_{\rm HII}$}
\newcommand{\RHII}{$R_{\rm HII}$}

\newcommand{\Halpha}{H$\alpha$}
\newcommand{\Hbeta}{H$\beta$}
\newcommand{\Hgamma}{H$\gamma$}
\newcommand{\Hdelta}{H$\delta$}
\newcommand{\Halphaw}{\Halpha\,$\lambda$6563}
\newcommand{\Hbetaw}{\Hbeta\,$\lambda$4861}
\newcommand{\Hgammaw}{H$\gamma$\,$\lambda$4340}
\newcommand{\Hdeltaw}{H$\delta$\,$\lambda$4101}
\newcommand{\Ha}{\Halpha}
\newcommand{\Hb}{\Hbeta}

\newcommand{\I}{\,{\sc i}}
\newcommand{\II}{\,{\sc ii}}
\newcommand{\III}{\,{\sc iii}}
\newcommand{\IV}{\,{\sc iv}}
\newcommand{\V}{\,{\sc v}}
\newcommand{\VI}{\,{\sc vi}}
\newcommand{\VII}{\,{\sc vii}}
\newcommand{\VIII}{\,{\sc viii}}

\newcommand{\HI}{H\I}
\newcommand{\HII}{H\II}
\newcommand{\HeI}{He\I}
\newcommand{\HeII}{He\II}

\newcommand{\CII}{[C\II]}
\newcommand{\CIIw}{\CII\,$\lambda$2325 (blend)}
\newcommand{\CIII}{[C\III]}
\newcommand{\CIIIw}{\CIII\,$\lambda$1908}
\newcommand{\CIIId}{C\III]}
\newcommand{\CIIIdw}{C\III]\,$\lambda\lambda$1907,1909}
\newcommand{\CIV}{C\IV}
\newcommand{\CIVw}{\CIV\,$\lambda$1549}
\newcommand{\OII}{[O\II]}
\newcommand{\OIIw}{\OII\,$\lambda$3727}
\newcommand{\OIIdw}{\OII\,$\lambda\lambda$3727,3729}
\newcommand{\OIII}{[O\III]}
\newcommand{\OIIIw}{\OIII\,$\lambda$5008}
\newcommand{\OIIIww}{\OIII\,$\lambda$4960,$\lambda$5008}
\newcommand{\OIIIdw}{\OIIIww}
\newcommand{\OIIIwa}{\OIII\,$\lambda$4363}
\newcommand{\OIIIwb}{O\III]\,$\lambda$1666}
\newcommand{\OIIIwc}{\OIII\,$\lambda$4960}
\newcommand{\NeIII}{[Ne\III]}
\newcommand{\NeIIIw}{\NeIII\,$\lambda$3869}
\newcommand{\NeIIIwb}{\NeIII\,$\lambda$3968}
\newcommand{\NII}{[N\II]}
\newcommand{\NIIw}{\NII\,$\lambda$6585}
\newcommand{\NIIww}{\NII\,$\lambda$6550,$\lambda$6585}
\newcommand{\SII}{[S\II]}
\newcommand{\SIIww}{\SII\,$\lambda$6718,$\lambda$6733}
\newcommand{\HeIw}{\HeI\,$\lambda$3889}
\newcommand{\HeIwa}{\HeI\,$\lambda$4473}
\newcommand{\HeIIw}{\HeII\,$\lambda$1640}
\newcommand{\HeIIwb}{\HeII\,$\lambda$4687}
\newcommand{\NIII}{N\III]}
\newcommand{\NIV}{N\IV]}
\newcommand{\NIIIw}{\NIII\,$\lambda$1748}
\newcommand{\NIVw}{\NIV\,$\lambda$1486}
\newcommand{\MgII}{Mg\II}
\newcommand{\MgIIw}{\MgII\,$\lambda$2800}

\newcommand{\Lyaw}{Ly$\alpha$\,$\lambda$1216}



\newcommand{\Om}{\Omega_{\rm M}}
\newcommand{\OL}{\Omega_\Lambda}

\newcommand{\etal}{et al.}

\newcommand{\citeps}{\citep}

\newcommand{\HST}{{\em HST}}
\newcommand{\SST}{{\em SST}}
\newcommand{\Hubble}{{\em Hubble}}
\newcommand{\Spitzer}{{\em Spitzer}}
\newcommand{\Chandra}{{\em Chandra}}
\newcommand{\JWST}{{\em JWST}}
\newcommand{\Planck}{{\em Planck}}

\newcommand{\Bradac}{{Brada\v{c}}}

\newcommand{\citepeg}[1]{\citep[e.g.,][]{#1}}

\newcommand{\range}[2]{\! \left[ _{#1} ^{#2} \right] \!}  

\newcommand{\grizli}{\textsc{grizli}}
\newcommand{\eazypy}{\textsc{eazypy}}
\newcommand{\msaexp}{\textsc{msaexp}}
\newcommand{\trilogy}{\textsc{trilogy}}
\newcommand{\bagpipes}{\textsc{bagpipes}}
\newcommand{\beagle}{\textsc{beagle}}
\newcommand{\photutils}{\textsc{photutils}}
\newcommand{\SEP}{\textsc{sep}}
\newcommand{\piXedfit}{\textsc{piXedfit}}
\newcommand{\pyneb}{\textsc{pyneb}}
\newcommand{\HIIC}{\textsc{hii-chi-mistry}}
\newcommand{\astropy}{\textsc{astropy}}
\newcommand{\astrodrizzle}{\textsc{astrodrizzle}}
\newcommand{\multinest}{\textsc{multinest}}
\newcommand{\cloudy}{\textsc{Cloudy}}
\newcommand{\jdaviz}{\textsc{Jdaviz}}

\renewcommand{\tt}[1]{\texttt{#1}}

\newcommand{\SE}{\tt{SourceExtractor}}

\newcommand{\PD}[1]{\textcolor{blue}{[PD: #1\;]}}


\begin{abstract}
Population III stars, the hypothetical first generation metal-free stars, have yet to be discovered.
Even after three years of successful JWST operations, studies have shown that most galaxies identified to date at $z > 5$ exhibit a metallicity floor of $Z\gtrsim2\%\,Z_{\odot}$, possibly due to unknown selection biases toward bright galaxies or rapid metal enrichment.
To address this question, we search for galaxies with low R3$=$\OIIIw/H$\beta$ ratios as part of the JWST Cycle-3 large treasury program, the Slitless Areal Pure-Parallel HIgh-Redshift Emission Survey (\texttt{SAPPHIRES}). 
Using deep NIRCam Wide-Field Slitless Spectrscopy (WFSS) data, we report the discovery of seven extremely metal-poor galaxy candidates in the \texttt{SAPPHIRES} Early Data Release field, with estimated $12+{\rm log(O/H)}<7.0$ at $z\sim5-7$, including {three} sources with $12+{\rm log(O/H)}<6.7$ ($Z<1\%\,Z_{\odot}$), significantly breaking the metallicity floor observed both locally and at high redshift.
These candidates appear extremely faint ($\sim28-30\,$ F200W AB mag) and low-mass (${\rm log}(M_{*}/M_{\odot})\sim6.8-7.8$), as expected from the mass-metallicity relation.
They also exhibit very blue UV slopes ({$-2.8\lesssim\beta\lesssim-1.7$}), likely due to low dust content $A_{V}\lesssim0.2\,{\rm mag}$ or young stellar ages $\sim5-20\,{\rm Myr}$.
Compared to galaxies at similar redshift, they appear exceptionally bursty in their star formation activity.
Our results highlight the power of NIRCam/WFSS in identifying extremely metal-poor galaxies, from just a single pointing, with more data to come in \texttt{SAPPHIRES}.
This underscores the potential of pure-parallel programs towards achieving JWST's primary science goal: discovering the first pristine stars and galaxies.
Deep JWST/NIRSpec follow-up observations will also be essential to confirm their nature and perform detailed chemical abundance analyses.

\end{abstract}


\section{Introduction} \label{sec:intro}
All elements heavier than hydrogen and helium (commonly referred to as metals) are believed to have been firstly produced by {the first generation of stars (Population {\sc iii} stars, or Pop {\sc iii}; e.g.,} \citealt[][]{Barkana2001,Bromm2013,Klessen2023}).
As a result, the first galaxies are expected to exhibit spectra characterized by strong Balmer lines (e.g., H$\alpha$, H$\beta$) and \HeII\ emission, but no metal lines \citep[e.g.,][]{Zackrisson2011}. 
Prior to JWST, searches for such galaxies were limited, with only a few possible candidates proposed based on strong \HeIIw\ emission, most notably, CR7 \citep{Sobral2015}.
However, subsequent studies challenged this interpretation, {detecting} the detection of \OIIIw\, suggesting that CR7 could instead be explained by an active galactic nuclei \citep[AGN; ][]{Bowler2017}.

JWST has revolutionized our ability to study galaxies at high redshift ($z>5$), owing to its unique combination of wavelength coverage, unprecedented sensitivity, and high spatial resolution in the rest-frame optical \citep[][]{Gardner06_jwst,Rigby22_jwst,Gardner2023}.
Several novel strategies have been proposed and implemented to search for the first generation of objects.
{These include medium-band photometric selection \citep[e.g.,][]{Nishigaki2023,Trussler2023,Fujimoto2025,Trussler2026}}, the detection of high-ionization helium recombination lines such as \HeII, either within galaxies \citep{Wang2024} or extended halos \citep{Maiolino2023,Maiolino2026}, and indirect Pop {\sc iii} signatures inferred from unusual carbon abundances \citep{DEugenio2023,Hsiao2024c}.
While these approaches offer promising insights, they are inherently indirect and often rely on sophisticated modelling.
Alternative interpretations, such as contributions from AGN, must also be considered and carefully modelled.
A more direct and interpretable method is to estimate the gas-phase metallicity of these galaxies.  

Although the quest for the ``Holy Grail'' of metal-free galaxies in the early Universe has begun in earnest, results from the first two years of JWST spectroscopy, consistently show that nearly all galaxies at $z>5$ exhibit metallicities $\gtrsim2\%$ solar \citep[e.g.,][]{Nakajima2023,Curti2024,Sanders2024}.
Remarkably, despite being observed over 13 billion years earlier, these galaxies display oxygen abundances comparable to those of metal-poor galaxies in the local Universe \citep[e.g.,][]{Isobe2022}.
The reason for this apparent ``metallicity floor'' remains unclear. 
One simple explanation is the rapid chemical enrichment in the early epochs \citep{Gutcke2022}. 
Another possibility is a selection bias introduced by conventional NIRCam color selection and NIRSpec follow-up strategies.
To date, only {two} remarkable stellar system has shown evidence of ultra low metallicity in the early universe: LAP-1 at $z=6.62$ with $Z\sim0.004\,Z_{\odot}$ inferred from {a low \OIIIw/H$\beta\sim0.7$; \citep{Vanzella2023,Nakajima2025}}, {and AMORE-6 at $z=5.73$ with $Z<0.002\,Z_{\odot}$ \citep{Morishita2025}}.
Notably, LAP-1 was first identified through its strong Lyman-$\alpha$ emission \citep{Vanzella2020}, yet remains undetected in NIRCam imaging {($M_{\rm UV}>-11.2$)}.
This discovery raises the possibility that other similar metal-poor systems exist, but are either undetected or only marginally detected in NIRCam wide-band observations, and thus have not been targeted for spectroscopic follow-up.

NIRCam Wide Field Slitless Spectroscopy (WFSS) enables the simultaneous acquisition of grism spectra for all galaxies within its $\sim10\,\rm{arcmin^2}$ field-of-view (FoV), thereby minimizing selection biases.
As such, NIRCam grism has proven to be an efficient tool for detecting emission-line galaxies, particularly via luminous rest-frame optical lines such as \OIIIw\ and H$\beta$ \citep[e.g.,][]{Sun2022,Sun2023,Sun_aspire,Kashino2023,Matthee2023,Wang2023,Oesch2023,Backhaus2024,Guo2024,Champagne2025}, which serve as key metallicity diagnostics.
NIRCam’s grism mode allows for the detection of \OIIIw\ and H$\beta$ at $z=5-7$ (using F356W) and $z=7-9$ (using F444W), enabling a systematic search for \OIIIw\ and H$\beta$ at high redshifts without relying on pre-selection from broadband colors.

In JWST Cycle 3, NIRCam WFSS became available in the pure parallel mode, enabling deep grism spectroscopy without primary observing time consumption. The Slitless Areal Pure-Parallel HIgh-Redshift Emission Survey (\texttt{SAPPHIRES}; GO 6434, PI: Egami) was designed for this purpose, representing one of the first pure-parallel JWST NIRCam WFSS programs.
\citet{Sun2025} presented the Early Data Release (EDR) from \texttt{SAPPHIRES},  which was attached to the GO 4750 program (PI: Nakajima, K.) for the NIRSpec MOS observations of the lensing cluster field MACS\,J0416.1–2403 (hereafter MACS0416), including LAP-1 \citep{Vanzella2023,Nakajima2025}.
In the parallel field, \texttt{SAPPHIRES} obtained deep NIRCam grism spectroscopy, with exposure times of approximately $\sim8\,$hrs in F356W and $\sim10\,$hrs in F444W.
These deep observations have already yielded significant results, including the spectroscopic identification of the most distant galaxy protocluster candidate to date at $z\sim8.47$ \citep{Fudamoto2025}.

In this work, we report the discovery and analysis of seven extremely metal-poor galaxies (EMPGs) with $12+{\rm log(O/H)}<7.0$ at $z\sim5-7$, identified in deep JWST grism observations from \texttt{SAPPHIRES}-EDR.
In Section \ref{Sec:odm}, we describe the observations, including the NIRCam imaging and grism data used in this study, as well as the procedures for emission line flux measurements.
We present the details of our SED fitting and EMPG candidates selection based on strong-line diagnostics in Section \ref{Sec:analysis}.
The results and discussion are presented in Section \ref{Sec:discussion}, followed by a summary of our main conclusions in Section \ref{sec:conclusion}.

Throughout this article, we adopt solar abundance ratios
\logOH\ = 8.69 \citep{Asplund2021}.
Where needed, we adopt the {\em Planck} 2018 flat \LCDM\ cosmology \citep{Planck18_cosmo}
with $H_0 = 67.7$ km s\inv\ Mpc\inv, $\Om = 0.31$, and $\OL = 0.69$.


\section{Observations, Data, and Measurement}
\label{Sec:odm}

This study makes use of data from the JWST Cycle 3 Treasury Survey, \texttt{SAPPHIRES} (GO‑6434; PIs: Egami, E., Fan, X., Sun, F., Wang, F.\ and Yang, J.).
\texttt{SAPPHIRES} is a pure-parallel JWST/NIRCam survey designed to obtain a total of {701.1} hours of observations, including {551} hours of exposure time, combining both imaging and grism spectroscopy.
In this work, we focus on the SAPPHIRES Early Data Release (EDR) in the MACS0416 parallel field. For details on the survey design, data reduction, photometric catalog, and spectral extraction, we refer readers to \citet{Sun2025}.
In the following section, we briefly summarize the EDR observations, describe the imaging and grism data used in this study, and outline our emission line measurements.

\subsection{EDR Observations} \label{sec:observation}
In brief, the \texttt{SAPPHIRES}‑EDR includes NIRCam imaging in eight broad-band and five medium-band filters at 0.6--4.8\,\micron\ over 16\,arcmin$^2$, along with F356W and F444W NIRCam WFSS data obtained in both Grism-C and Grism-R dispersion directions within the same pointings.
The F356W and F444W grism exposures have integration times of approximately 4.7 hours and 7.1 hours, respectively, per dispersion direction. These deep integrations yield $5\,\sigma$ line sensitivities of $\sim0.5-1\times10^{-18}\,{\rm erg\,s^{-1}\,cm^{-2}}$, making \texttt{SAPPHIRES}-EDR one of the deepest NIRCam grism datasets to date.
For the imaging component, \texttt{SAPPHIRES}-EDR achieves $5\,\sigma$ depth of $\sim28.5-29.0\,{\rm AB mag}$ depending on the filter. 
We refer readers to Tables 1 and 2 in \citet{Sun2025} for detailed exposure times across all bands and observing modes.

Although the EDR field is parallel to the Frontier Field cluster MACS\,J0416.1--2403, the pointing lies at a relatively large angular separation ($\gtrsim4^{\prime}$) from the galaxy cluster. 
Therefore, gravitational lensing effects are negligible for the purpose of this study ($\mu\lesssim1.1$; according to the lens model of \citealt{Richard2021}){, and no correction for lensing is applied to the masses and SFRs.}

\subsection{Imaging and Grism Data} \label{sec:data}

The \texttt{SAPPHIRES} EDR NIRCam imaging data were processed using the standard \textsc{jwst} pipeline, with several customized steps to correct for known instrumental artifacts.
{\citet{Sun2025} used \texttt{photutils} \citep{photutils}} to identify sources and perform aperture photometry.
Photometry was measured using four circular apertures ($r=0\farcs10$, $0\farcs15$, $0\farcs25$, and $0\farcs30$) and two Kron apertures ($K=2.5$ and $1.2$), with point-source aperture loss corrected, resulting in a comprehensive photometric catalog, which is publicly available on the \texttt{SAPPHIRES} website\footnote{\url{https://jwst-sapphires.github.io}}.

\texttt{SAPPHIRES} EDR WFSS data were also reduced using the \textsc{jwst} pipeline, incorporating a series of tailored steps following the methods described in \citet{Sun2022,Sun2023}.
These procedures include flat-fielding, background subtraction, bad pixel masking, and continuum subtraction. Grism spectra were extracted for all sources with magnitudes brighter than $29.8\,$AB mag in the F356W band or $29.6\,$AB mag in the F444W band.
1D spectra were extracted from the 2D spectra both with and without continuum subtraction{, where the continuum is dominated by the residual sky background and sources overlapping along the dispersion direction,} using both a boxcar aperture (height = $0\farcs31$) and an optimal extraction method \citep{horne86} based on the source profile.
Note that contamination from neighboring sources was not taken into account in the extraction process.

The spectroscopic redshifts are estimated based on several factors.
First, the observed, continuum-subtracted spectrum is cross-matched with a set of template spectra, optimizing for different combinations of emission lines.
The resulting $\chi^2(z)$ distribution is then compared to the photometric redshift probability distribution, and a confidence level is assigned to each spectroscopic redshift accordingly, also taking into account the number of confidently detected emission lines.
In addition, the best-fit redshift solution for each galaxy is visually inspected to ensure reliability.
Details of the templates, fitting algorithm, and redshift estimation procedure can be found in \citet{LinX2025a} and \citet{Sun2025}.

\subsection{Line Measurements} \label{sec:measurement}

In this process, we begin with the \texttt{SAPPHIRES} EDR catalog, which includes spectroscopic redshifts for 1,060 galaxies spanning $0\lesssim z\lesssim8.5$.
We then use the 1D spectra extracted via the optimal extraction method to measure the fluxes of the \OIIIw\ and H$\beta$ emission lines.
For each line fitting, we first apply a spline fit to the continuum, masking out regions ($\pm0.03\,{\rm \mu m}$; $\sim2500\,{\rm km/s}$ at $z=6$) around the emission lines.
The emission lines are modelled using a single Gaussian profile convolved with the line spread function (LSF) appropriate for the NIRCam R and C grism channels (Sun et al.\ in prep.).
All fitting procedures are performed independently for the R and C grism spectra {to avoid possible contamination from nearby sources}.
For the \OIIIw\ line, the central wavelength, line width, and flux are treated as free parameters, with the redshift fixed to the values provided in the EDR catalog.
For the \OIIIwc\ and H$\beta$ lines, the line widths are fixed to match that of \OIIIw, {
while the central wavelengths are allowed to vary within 0.05$\,\mu$m.
All fits remain within a reasonable range of the expected wavelengths.
If $z\lesssim6.6$ (such that H$\alpha$ falls within the F444W coverage) and H$\alpha$ does not fall in the chip gap, H$\alpha$ is also measured.}
If the fitting {has the SNR too low (${\rm SNR}<3$)}, we then allow the line widths to be also free parameters.
{We also estimate the equivalent width (EW) of the emission line measured.
Since we do not detect the stellar continuum in the grism spectra, we estimate the stellar continuum from the photometry (see also \S\ref{sec:SED}).
}

\begin{deluxetable*}{lccccccc}
\tablecaption{\label{tab:phot}Photometry (in nJy) of extremely metal-poor galaxy candidates, measured in $r=0.1''$ aperture. $1\,$nJy = $31.4\,$AB mag.}
\tablewidth{\columnwidth}
\tablehead{
\colhead{ID} &
\colhead{1449} &
\colhead{7695} &
\colhead{9770} &
\colhead{20130} &
\colhead{21782} &
\colhead{24804} &
\colhead{37141}
}
\startdata
RA (J2000)   & 63.96689  & 63.95838  & 64.00681  & 63.95847  & 63.99127  & 63.97931  & 63.92739 \\
DEC (J2000)  & $-24.19149$ & $-24.16981$ & $-24.16483$ & $-24.14112$ & $-24.13627$ & $-24.12402$ & $-24.15730$ \\ \hline
F070W   & $ -1.4 \pm 1.2 $ & $ 3.0 \pm 1.0 $ & $ -1.6 \pm 1.3 $ & $ 1.4 \pm 1.4 $ & $ 2.0 \pm 1.4 $ & $ -0.8 \pm 1.2 $ & $ 4.8 \pm 1.4 $ \\
F090W   & $ 6.2 \pm 1.4 $ & $ 26.3 \pm 1.3 $ & $ 6.5 \pm 1.4 $ & $ 6.2 \pm 1.5 $ & $ 5.5 \pm 1.1 $ & $ 3.7 \pm 1.4 $ & $ 27.9 \pm 1.7 $ \\
F115W   & $ 7.6 \pm 1.3 $ & $ 25.7 \pm 1.3 $ & $ 11.6 \pm 1.3 $ & $ 20.4 \pm 1.5 $ & $ 6.1 \pm 1.5 $ & $ 3.3 \pm 1.3 $ & $ 24.5 \pm 1.5 $ \\
F140M   & ... & $ 19.8 \pm 1.4 $ & ... & ... & $ 10.2 \pm 1.5 $ & ... & ... \\
F150W   & $ 11.3 \pm 1.0 $ & $ 19.7 \pm 1.0 $ & $ 9.7 \pm 1.1 $ & $ 17.6 \pm 1.2 $ & $ 6.9 \pm 1.2 $ & $ 2.4 \pm 1.1 $ & $ 23.6 \pm 1.2 $ \\
F182M   & $ 9.1 \pm 1.5 $ & $ 19.3 \pm 1.2 $ & $ 6.2 \pm 1.5 $ & $ 17.6 \pm 1.6 $ & $ 6.3 \pm 1.7 $ & $ 4.8 \pm 1.5 $ & $ 21.8 \pm 1.7 $ \\
F200W   & $ 10.4 \pm 1.0 $ & $ 16.9 \pm 0.8 $ & $ 7.9 \pm 1.0 $ & $ 19.1 \pm 1.0 $ & $ 5.4 \pm 1.1 $ & $ 2.2 \pm 1.0 $ & $ 23.7 \pm 1.2 $ \\
F210M   & $ 9.5 \pm 1.8 $ & $ 19.3 \pm 1.5 $ & $ 10.1 \pm 1.9 $ & $ 17.4 \pm 2.0 $ & $ 10.5 \pm 2.0 $ & $ 2.2 \pm 1.8 $ & $ 23.5 \pm 2.1 $ \\
F277W   & $ 8.2 \pm 0.8 $ & $ 13.3 \pm 0.7 $ & $ 9.5 \pm 0.9 $ & $ 15.7 \pm 0.9 $ & $ 5.8 \pm 0.7 $ & $ 4.1 \pm 0.9 $ & $ 19.0 \pm 1.0 $ \\
F335M   & ... & $ 22.2 \pm 1.4 $ & ... & ... & $ 19.0 \pm 1.7 $ & ... & ... \\
F356W   & $ 14.5 \pm 0.9 $ & $ 15.8 \pm 0.8 $ & $ 14.0 \pm 0.9 $ & $ 28.6 \pm 1.4 $ & $ 10.5 \pm 0.6 $ & $ 4.8 \pm 0.9 $ & $ 26.4 \pm 1.3 $ \\
F410M   & $ 7.9 \pm 1.6 $ & $ 7.8 \pm 1.2 $ & $ 2.9 \pm 1.5 $ & $ 34.2 \pm 1.7 $ & $ 3.2 \pm 1.1 $ & $ 4.0 \pm 1.6 $ & $ 16.7 \pm 2.0 $ \\
F444W   & $ 10.8 \pm 1.2 $ & $ 13.1 \pm 0.7 $ & $ 11.9 \pm 1.3 $ & $ 20.9 \pm 1.3 $ & $ 8.3 \pm 0.8 $ & $ 2.1 \pm 1.3 $ & $ 18.1 \pm 1.5 $ \\
\enddata
\end{deluxetable*}


\section{Analysis}
\label{Sec:analysis}

\subsection{Selection Criteria and Strong-Line Diagnostics}
\label{Sec:strongline}
Traditionally, deriving gas-phase metallicities requires knowledge of key physical conditions, such as the electron temperature and electron density. 
However, temperature-sensitive auroral lines (e.g., \OIIIwa\ and \OIIIwb) are typically faint, and thus difficult to detect, particularly in EMPGs at high redshifts, which are expected to be low-mass and faint.
Given these limitations, we instead adopt the R3=\OIIIw/H$\beta$ ratio, as a strong-line diagnostic of metallicity. 
These two emission lines are among the brightest in the rest-frame optical, and their ratio provides a practical metallicity indicator in the absence of auroral lines.
We begin by selecting galaxies in the redshift ranges $5<z<7$ (F356W) and $7<z<9$ (F444W), from the spectroscopic redshift catalog of 1060 sources reported by \citet{Sun2025}. 
We fit both the \OIIIw\ and H$\beta$ emission lines in the F356W and F444W filters for about {163 and 34} galaxies in the $5<z<7$ and $7<z<9$ bins, respectively. 
The R3 ratio is then computed based on the fitted line fluxes to identify candidates for extremely metal-poor galaxies.
We subsequently select galaxies with signal-to-noise ratios (SNR) greater than 3 for both \OIIIw\ and H$\beta$, {applying this criterion independently to the R and C grism spectra}.

\begin{figure}
\centering
\includegraphics[width=\columnwidth]{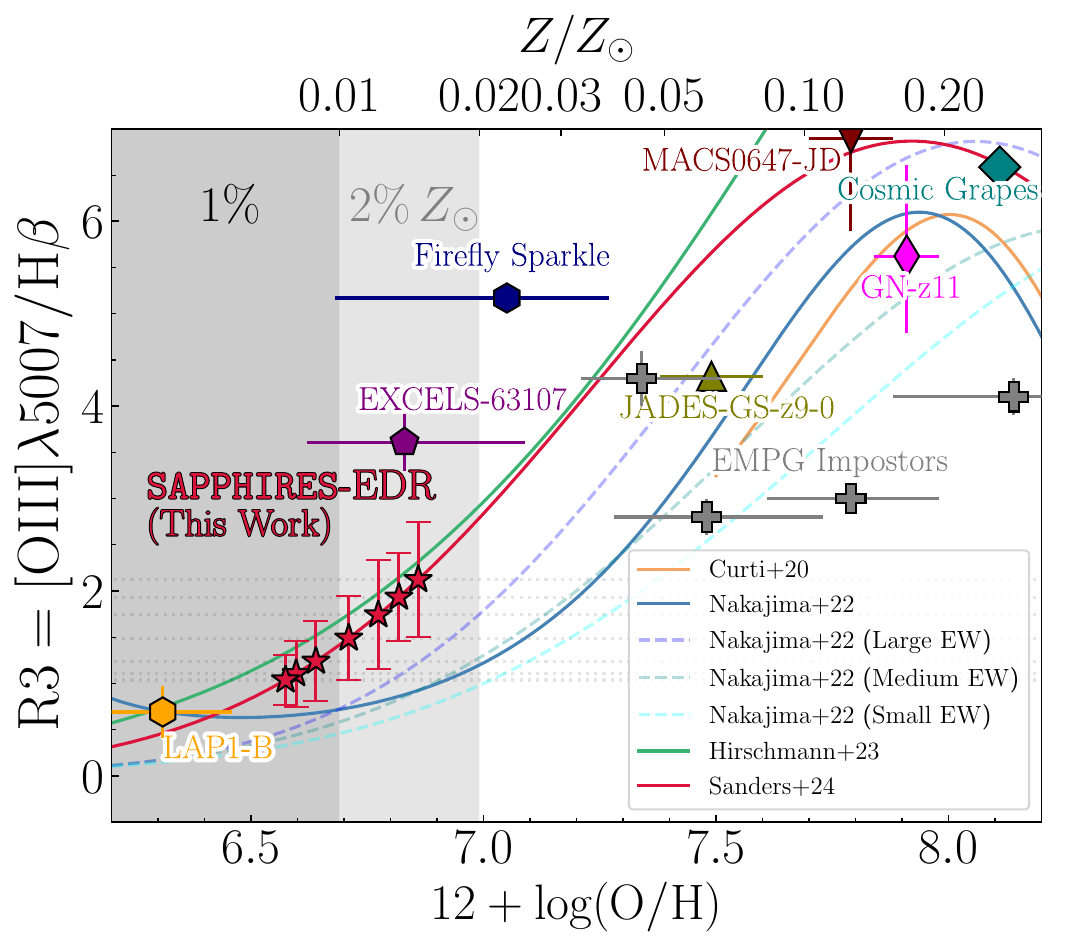}
\caption{Strong-line calibrations (R3 vs metallicity) assumed \citep[the crimson line;][]{Sanders2024} for the EMPG candidates (the crimson stars).
We present some other strong-line relations, including \citealt{Curti2020} (the orange line), \citealt{Hirschmann2023} (the green line), \citealt{Nakajima2022} (the blue lines) and the relation of different EW of H$\beta$ (the blue dashed lines).
We also compare the strong-line relations with several well-studied high-redshift galaxies with direct metallicities: MACS0647$-$JD \citep[the upside-down maroon triangle;][]{Hsiao2024b}, JADES-GS-z9-0 \citep[the olive triangle;][]{Curti2024b}, the Cosmic Grapes \citep[the teal diamond;][]{Fujimoto2024}, the Firefly Sparkle \citep[the navy hexagon;][]{Mowla2024}, GN-z11 \citep[the thin pink diamond;][]{Alvarez2024}, EXCELS-63107 \citep[the purple pentagon;][]{Cullen2025}, CANUCS-A370-z8-LAE \citep[the green octagon;][]{Willott2025}, {LAP1-B \citep[the orange hexagon;][]{Vanzella2023,Nakajima2025}, and EMPG impostors \citep[the gray pluses;][]{Hsiao2026b}}.
Gray shaded-regions indicate the metallicity floor of $2\%$ (lighter) and $1\%$ (darker) solar metallicity and the gray dotted horizontal lines highlight the R3 of our EMPG candidates.
}
\label{fig:R3}
\end{figure}

The goal of this study is to identify galaxies that break the observed ``metallicity floor'', typically defined as $Z\lesssim2\%Z\,_{\odot}$ or $12+{\rm log(O/H)}\lesssim7.0$.
This corresponds approximately to ${\rm R3}\lesssim3$ \citep[${\rm log(R3)}=0.834-0.072x-0.453x^{2}$, $x=4+{\rm log(O/H)};$][]{Sanders2024}.
Therefore, we next filter the sample to include only galaxies with ${\rm R3}<3$. 
Each candidate is manually and visually inspected, in both 1D and secure detection in 2D spectra, resulting in a final sample of seven EMPG candidates, all of which lie in the redshift of $5<z<7$.
In the end, seven candidates stand out as potential EMPGs.
The spectra of these EMPG candidates are presented in Figures~\ref{fig:spec1} and \ref{fig:spec2}.
The coordinates and photometry are presented in {Table} \ref{tab:phot}.
The spectroscopic redshifts and the measured line fluxes for H$\beta$, \OIIIw, and H$\alpha$, {and their rest-frame EWs} are summarized in Table \ref{tab:spec}.
We also construct a stacked spectrum (see \S\ref{Sec:stack}.)
None of the seven sources exhibit detectable auroral lines or density-sensitive doublets (e.g., \OIIdw).
Therefore, we estimate their metallicities using the strong-line diagnostics of \citet{Sanders2024}, which is constrained from high-redshift galaxies with JWST.
The metallicities converted through R3 using different strong-line calibrations are illustrated in Figure \ref{fig:R3}.

\subsection{Spectral Energy Distribution (SED) Fitting} \label{sec:SED}
We perform spectral energy distribution (SED) fitting for our sample of seven extremely metal-poor galaxy candidates using \textsc{Bagpipes} \citep{carnall18} to estimate their physical properties, including stellar mass. 
The modelling methodology, including adopted priors, assumptions, and model choices, is detailed in \citet{Hsiao2023a,Hsiao2024a} and the \texttt{SAPPHIRES} EDR paper \citep{Sun2025}, unless stated otherwise.
Briefly, to account for binary stellar evolution, we adopt the Binary Population and Spectral Synthesis (BPASS) v2.2.1 models \citep{Stanway2018}.
Nebular emission is included by reprocessing the stellar templates through the photoionization code Cloudy \citep{Ferland2017}.
We assume a delayed-$\tau$ star formation history {(parameterized as ${\rm SFR}(t)\propto t\times {\rm exp}(-t/\tau)$, where $t$ is the age and $\tau$ is a parameter that controls the duration of the star formation)}, adopt the \citet{Calzetti2000} dust attenuation law (with $0<A_{V}<8$ and a fixed $\eta=2$), and a \citet{Kroupa1993} initial mass function. 
{The stellar metallicity is allowed to vary between $0.5\%\,Z_{\odot}$ to $2\,Z_{\odot}$, assuming a prior uniform in logarithmic (base-10) space.
We also tested fixing the stellar metallicity to the gas-phase metallicity derived from the grism spectroscopic data, but the results did not change significantly.}
A 5$\%$ photometric error floor is imposed to account for potential systematic uncertainties due to calibration.

In addition to the physical properties catalog released in the EDR \citep[based on Kron aperture photometry; ][]{Sun2025}, we also fit photometry measured in multiple circular apertures ($r=0\farcs10$, $0\farcs15$, $0\farcs25$, and $0\farcs30$; see also \S\ref{sec:data}).
Because galaxies in our sample are compact in morphology, the results derived from the $r=0\farcs10$ aperture are adopted as our fiducial values throughout this study.
For four of the EMPG candidates, the derived physical properties are consistent across all four circular apertures. However, for the three faintest EMPG candidates: 9770, 21782, and 24804, the results become inconsistent when using photometry from larger apertures (i.e., $r=0\farcs25$, and $0\farcs30$), likely caused by the increased photometric error with those apertures.

The physical properties estimated from the SED fitting, including stellar masses, star formation rates (SFR), dust attenuation $A_V$, mass-weighted stellar ages, UV continuum slopes $\beta$, and metallicities of candidates, are presented in Table \ref{tab:SED}.
We note that for 24804, there are only detections {(SNR$>$3)} in three bands, making the SED fitting results highly unconstrained.

\begin{deluxetable*}{lcccccc}
\tablecaption{\label{tab:spec}
Spectroscopic redshifts, emission line fluxes, R3 ratios, and the metallicities of EMPG candidates.}
\tablewidth{\columnwidth}
\tablehead{
\colhead{ID} &
\colhead{$z_{\rm spec}$} &
\colhead{H$\beta$ (EW(H$\beta$))} &
\colhead{\OIIIw\ (EW(\OIII))} &
\colhead{H$\alpha$} &
\colhead{R3} &
\colhead{$12+{\rm log(O/H)}^{a}$} 
\\
\colhead{} &
\colhead{} &
\colhead{$10^{-19}\,{\rm erg\,s^{-1}\,cm^{-2}}$ (${\rm \AA}$)} &
\colhead{$10^{-19}\,{\rm erg\,s^{-1}\,cm^{-2}}$ (${\rm \AA}$)} &
\colhead{$10^{-19}\,{\rm erg\,s^{-1}\,cm^{-2}}$ (${\rm \AA}$)} &
\colhead{} & 
\colhead{} 
}
\startdata
1449 & 5.92 & $ 9.3 \pm 2.4 \ ( 310\pm  80)$ & $ 13.9 \pm 2.2 \ ( 510\pm  80)$ & $ <7.2\ ( <90)$ & $ 1.49 \pm 0.45 $ & $ 6.71 ^{+ 0.11 }_{- 0.14 }$ \\
7695 & 5.80 & $ 8.9 \pm 2.4 \ ( 300\pm  80)$ & $ 9.8 \pm 1.8 \ ( 360\pm  70)$ & $ 10.5 \pm 2.9 \ ( 120\pm  30)$ & $ 1.10 \pm 0.36 $ & $ 6.60 ^{+ 0.11 }_{- 0.14 }$ \\
9770 & 6.29 & $ 7.7 \pm 1.6 \ ( 250\pm  50)$ & $ 8.0 \pm 1.3 \ ( 280\pm  50)$ & $ <6.3 \ (<60)$ & $ 1.04 \pm 0.27 $ & $ 6.57 ^{+ 0.09 }_{- 0.14 }$ \\
20130 & 6.79 & $ 7.9 \pm 1.7 \ ( 240\pm  50)$ & $ 15.2 \pm 1.8 \ ( 490\pm  60)$ & \ldots & $ 1.93 \pm 0.48 $ & $ 6.82 ^{+ 0.10 }_{- 0.14 }$ \\
21782 & 5.81 & $ 6.4 \pm 2.0 \ ( 220\pm  70)$ & $ 7.9 \pm 1.3 \ ( 290\pm  50)$ & $ 5.6 \pm 1.4 \ (  60\pm  20)$ & $ 1.24 \pm 0.43 $ & $ 6.64 ^{+ 0.12 }_{- 0.14 }$ \\
24804 & 6.67 & $ 3.8 \pm 1.0 \ ( 120\pm  30)$ & $ 6.7 \pm 1.3 \ ( 220\pm  40)$ & \ldots & $ 1.74 \pm 0.59 $ & $ 6.77 ^{+ 0.13 }_{- 0.14 }$ \\
37141 & 5.76 & $ 10.0 \pm 2.4 \ ( 350\pm  80)$ & $ 21.2 \pm 3.5 \ ( 790\pm 130)$ & \ldots & $ 2.12 \pm 0.62 $ & $ 6.86 ^{+ 0.12 }_{- 0.14 }$ \\
\enddata
\tablenotetext{a}{Converted from R3 using strong-line relation from \citet{Sanders2024}.}
{}

\end{deluxetable*}

\begin{deluxetable*}{lcccccccc}
\tablecaption{\label{tab:SED}Physical properties and the UV continuum slope of EMPG candidates.}
\tablewidth{\columnwidth}
\tablehead{
\colhead{ID} &
\colhead{$z_{\rm spec}$} &
\colhead{Stellar Mass} &
\colhead{SFR$_{\rm SED}^{a}$} &
\colhead{SFR$_{\rm H\alpha}$} &
\colhead{$A_V$} &
\colhead{Mass-weighted Age} &
\colhead{$\beta^{b}$} &
\colhead{{$\tau^{c}$}}
\\
\colhead{} &
\colhead{} &
\colhead{log($M_{*}/M_\odot$)} &
\colhead{$M_{*}$/yr} &
\colhead{$M_{*}$/yr} &
\colhead{mag} &
\colhead{Myr} &
\colhead{}&
\colhead{{Myr}}
}
\startdata
1449 & $5.92$ & $7.44^{+0.19}_{-0.18}$ & $1.27^{+0.57}_{-0.43}$ & \ldots &$0.22^{+0.08}_{-0.09}$ & $ 14^{+ 16}_{-  8}$ & {$-1.7\pm0.6$} & {$5\pm3$} \\
7695 & $5.80$ & $7.20^{+0.11}_{-0.15}$ & $1.42^{+0.09}_{-0.29}$ & $1.29\pm0.36$ &$0.01^{+0.01}_{-0.01}$ & $  5^{+  3}_{-  3}$ & {$-2.8\pm0.1$} & {$5^{+4}_{-3}$} \\
9770 & $6.29$ & $7.14^{+0.32}_{-0.37}$ & $0.68^{+0.13}_{-0.13}$ & \ldots &$0.05^{+0.04}_{-0.03}$ & $ 11^{+ 21}_{-  8}$ & {$-2.7\pm0.2$} & {$5\pm3$} \\
20130 & $6.79$ & $7.76^{+0.19}_{-0.18}$ & $1.72^{+0.55}_{-0.32}$ & \ldots &$0.08^{+0.06}_{-0.04}$ & $ 23^{+ 22}_{- 11}$ & {$-2.1\pm0.2$} & {$5\pm3$} \\
21782 & $5.81$ & $6.81^{+0.23}_{-0.22}$ & $0.55^{+0.20}_{-0.16}$ & $0.69\pm0.17$ &$0.09^{+0.08}_{-0.05}$ & $  5^{+  4}_{-  3}$ & {$-2.3\pm0.4$} & {$6^{+3}_{-4}$}  \\
24804 & $6.67$ & $7.49^{+0.20}_{-0.27}$ & $0.24^{+0.16}_{-0.07}$ & \ldots &$0.14^{+0.15}_{-0.10}$ & $105^{+105}_{- 68}$ & \ldots & {$5\pm3$} \\
37141 & $5.76$ & $7.64^{+0.12}_{-0.14}$ & $1.91^{+0.42}_{-0.35}$ & \ldots &$0.06^{+0.04}_{-0.04}$ & $ 15^{+ 11}_{-  6}$ & {$-2.1\pm0.1$} & {$5^{+3}_{-4}$} \\
\enddata
\tablenotetext{a}{Averaged over the last 10$\,$Myr before epoch of observation.}
\tablenotetext{b}{UV continuum slope ($f_\lambda \propto \lambda^\beta$) estimated using {F115, F150W, and F200W.}}
{\tablenotetext{c}{The parameter controlling the duration of the star formation in the assumed delayed-$\tau$ SFH.}}
{}

\end{deluxetable*}



\begin{figure*}
    \centering
    \begin{minipage}[b]{0.63\textwidth}
        \includegraphics[width=\textwidth]{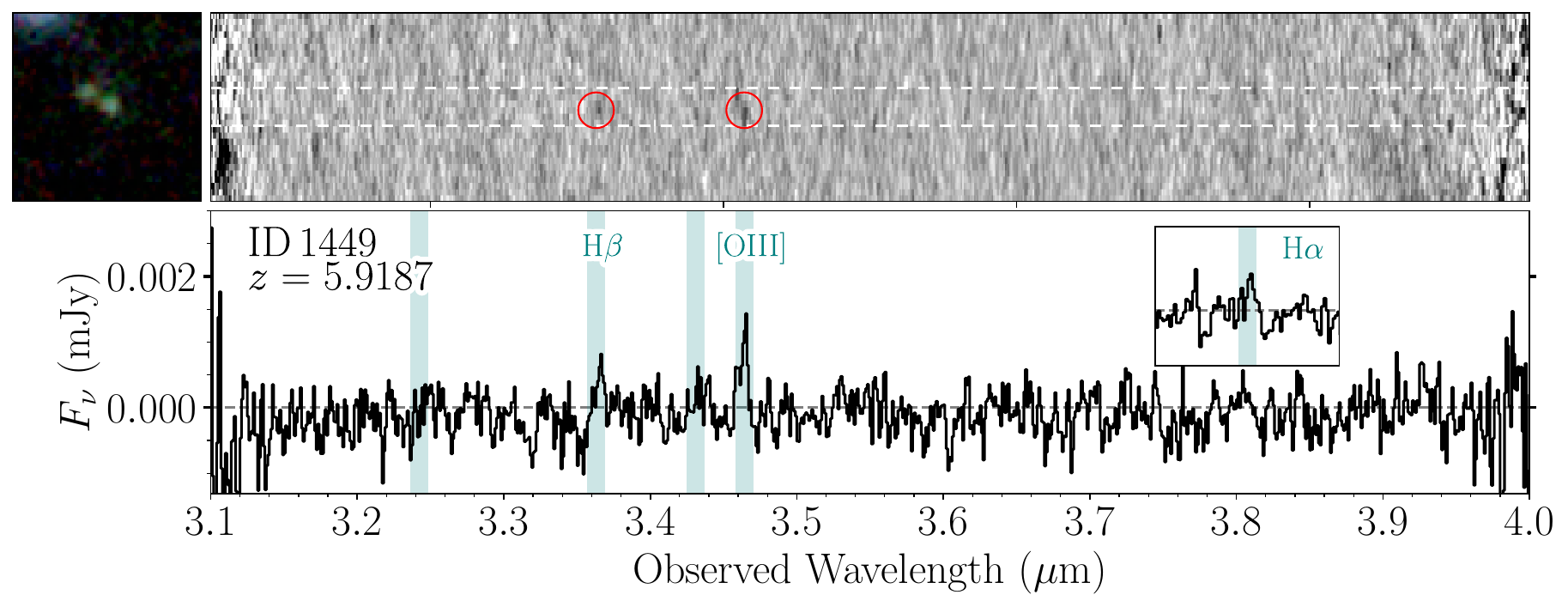}
    \end{minipage}
    \begin{minipage}[b]{0.35\textwidth}
        \includegraphics[width=\textwidth]{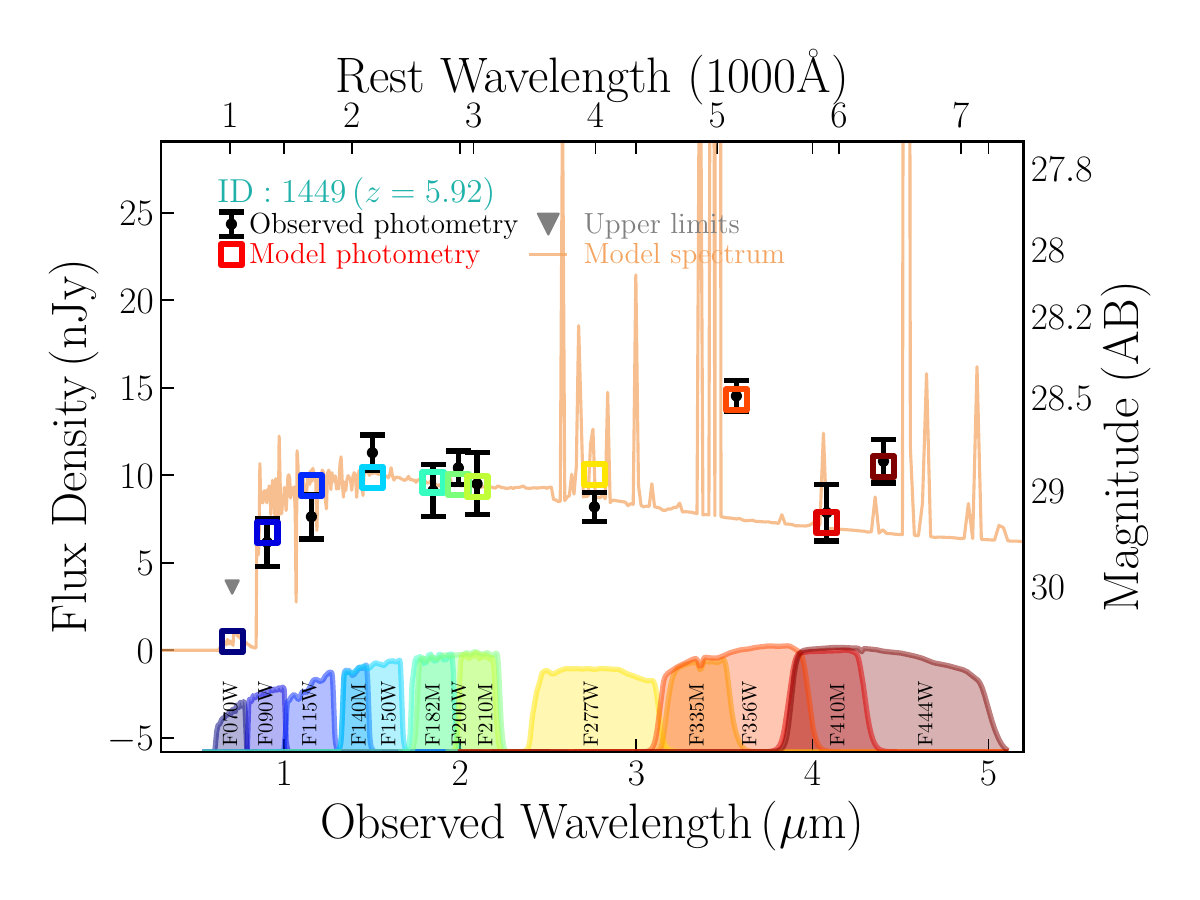}
    \end{minipage}
    \vspace{0.3cm} 
    \centering

    \begin{minipage}[b]{0.63\textwidth}
        \includegraphics[width=\textwidth]{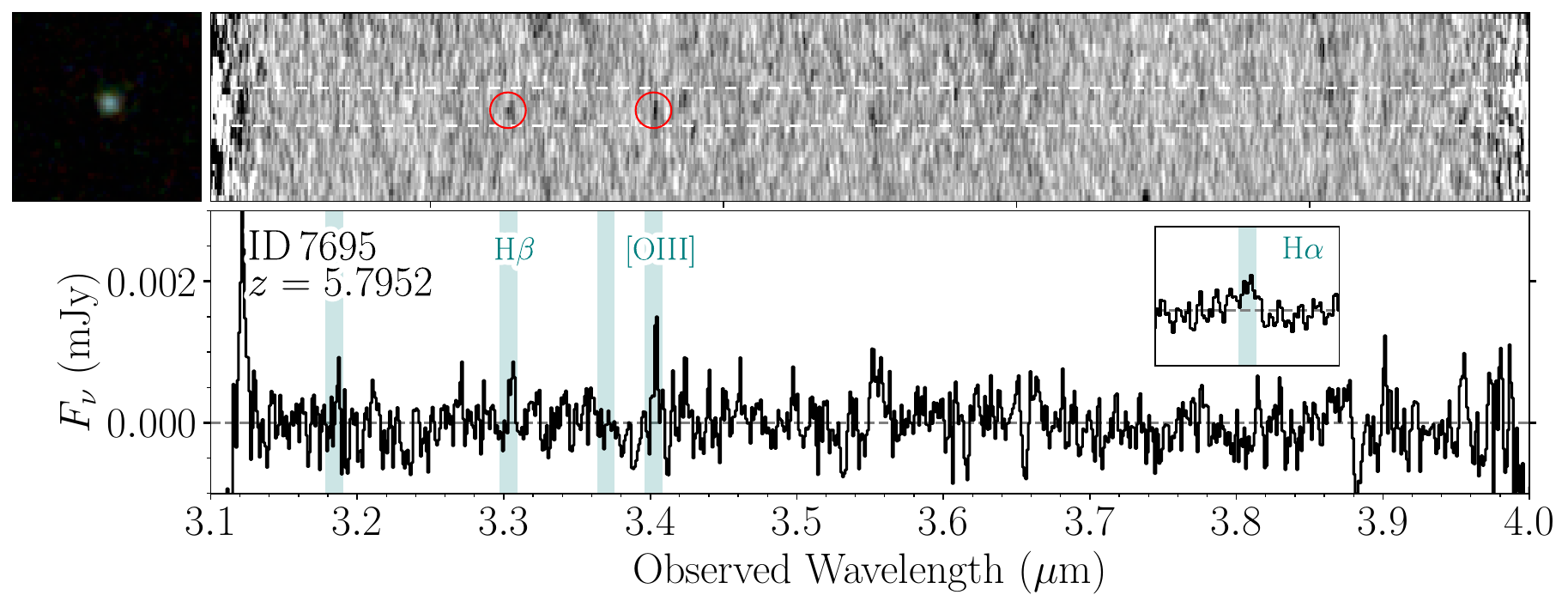}
    \end{minipage}
    \begin{minipage}[b]{0.35\textwidth}
        \includegraphics[width=\textwidth]{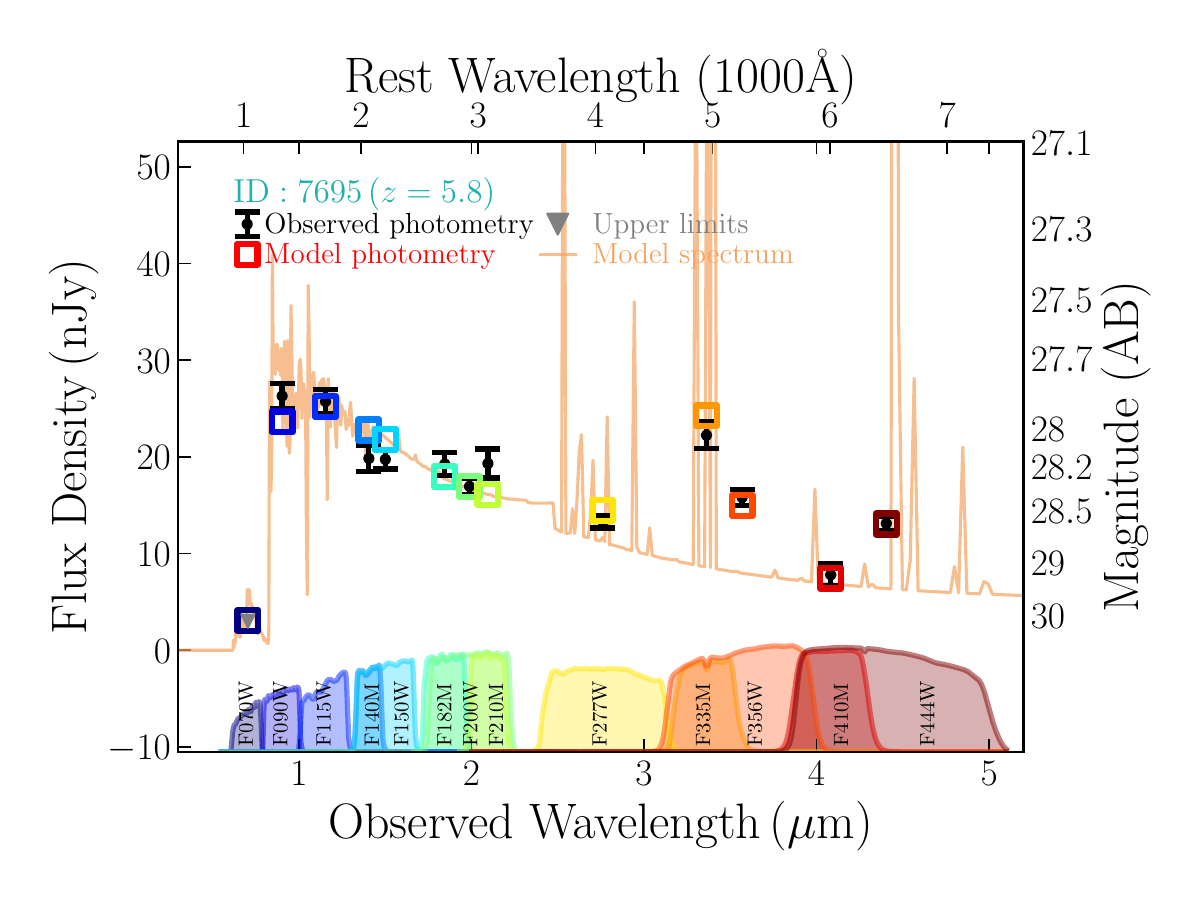}
    \end{minipage}
    \vspace{0.3cm} 

    \begin{minipage}[b]{0.63\textwidth}
        \includegraphics[width=\textwidth]{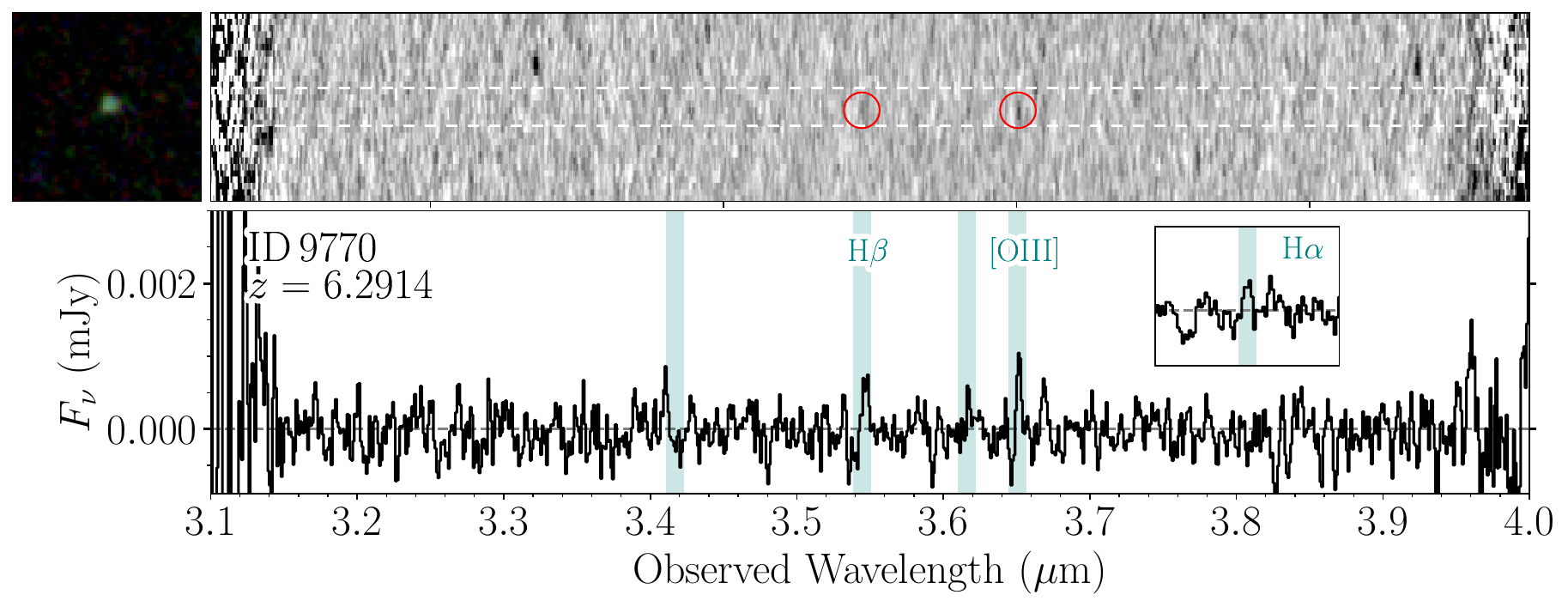}
    \end{minipage}
    \begin{minipage}[b]{0.35\textwidth}
        \includegraphics[width=\textwidth]{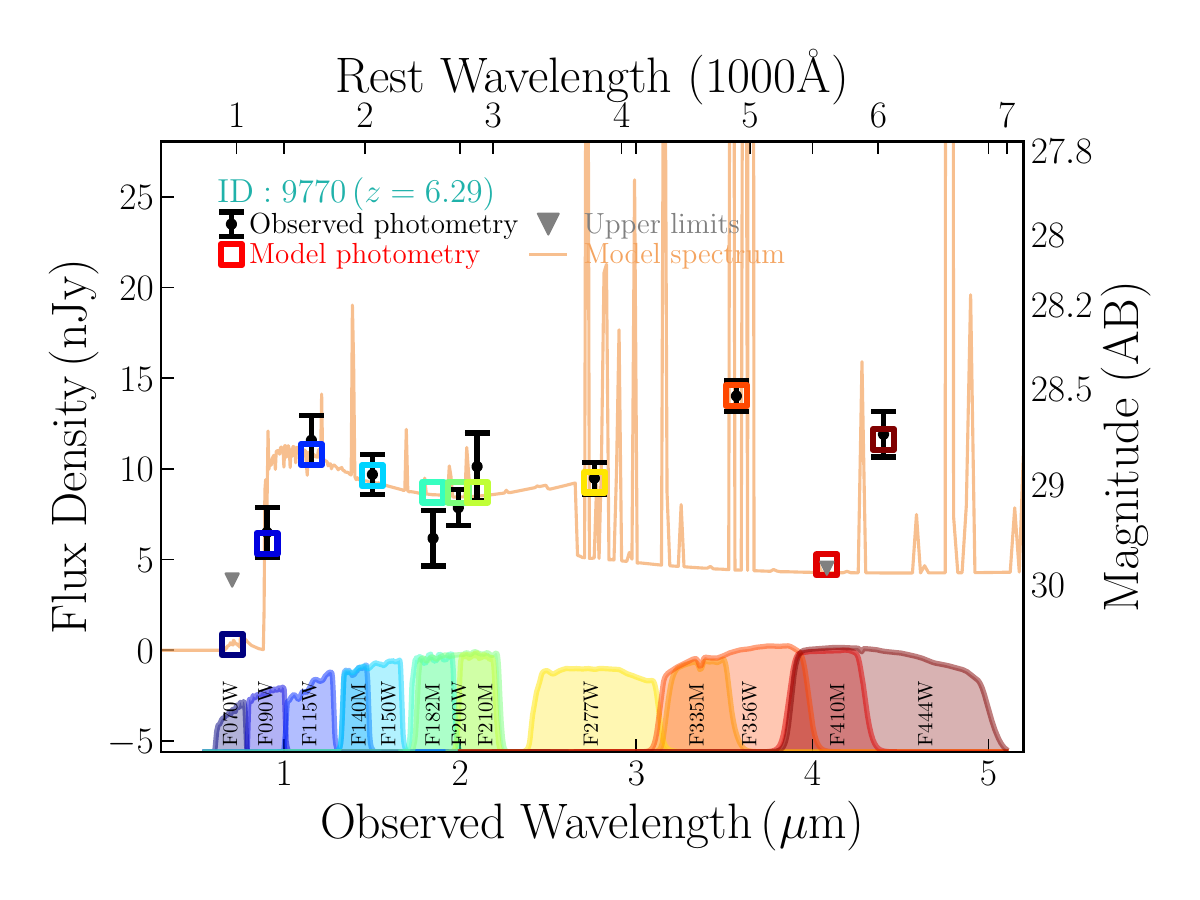}
    \end{minipage}
    \vspace{0.3cm} 
\caption{Left: The color images, 2D spectra, and extracted 1D spectra of each EMPG candidate.
The color images are made from Red: F444W- Green: F356W- Blue: F277W band images, and the images size are $3''\times3''$.
The red circles mark the locations of H$\beta$ and \OIIIw\ in the 2D spectra.
The black curves show the observed 1D spectrum, and the wavelengths of H$\beta$ and \OIIIw\ are highlighted in teal shaded regions.
Right: SED fittings of each EMPG candidate.
Black dots with error bars are photometry observed with JWST and the model photometry is indicated by {colored} squares.
Orange curves represent the best-fits in \textsc{Bagpipes} and the grey triangles are upper limits ($3\sigma$) for the photometry.
\label{fig:spec1}}
\end{figure*}

\begin{figure*}
    \centering
    \begin{minipage}[b]{0.63\textwidth}
        \includegraphics[width=\textwidth]{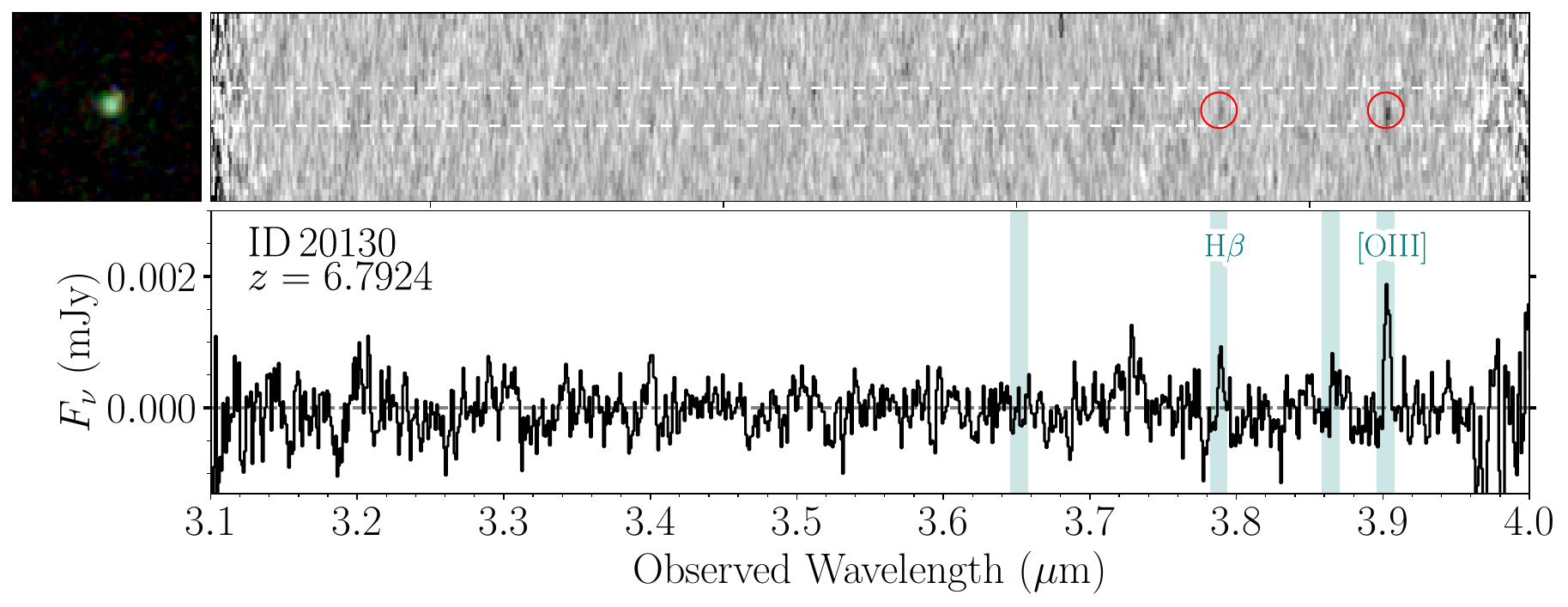}
    \end{minipage}
    \begin{minipage}[b]{0.35\textwidth}
        \includegraphics[width=\textwidth]{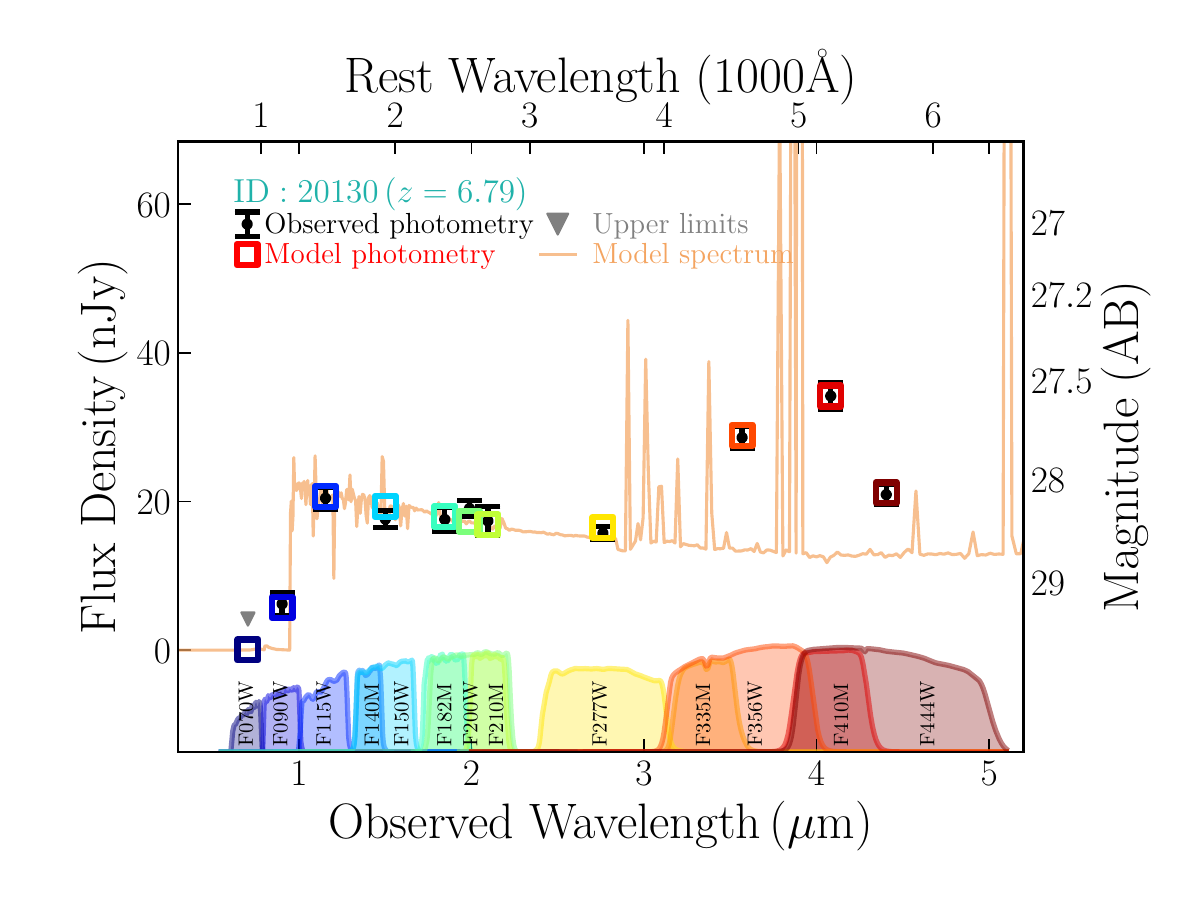}
    \end{minipage}
    \vspace{0.3cm} 
    \centering
    \begin{minipage}[b]{0.63\textwidth}
        \includegraphics[width=\textwidth]{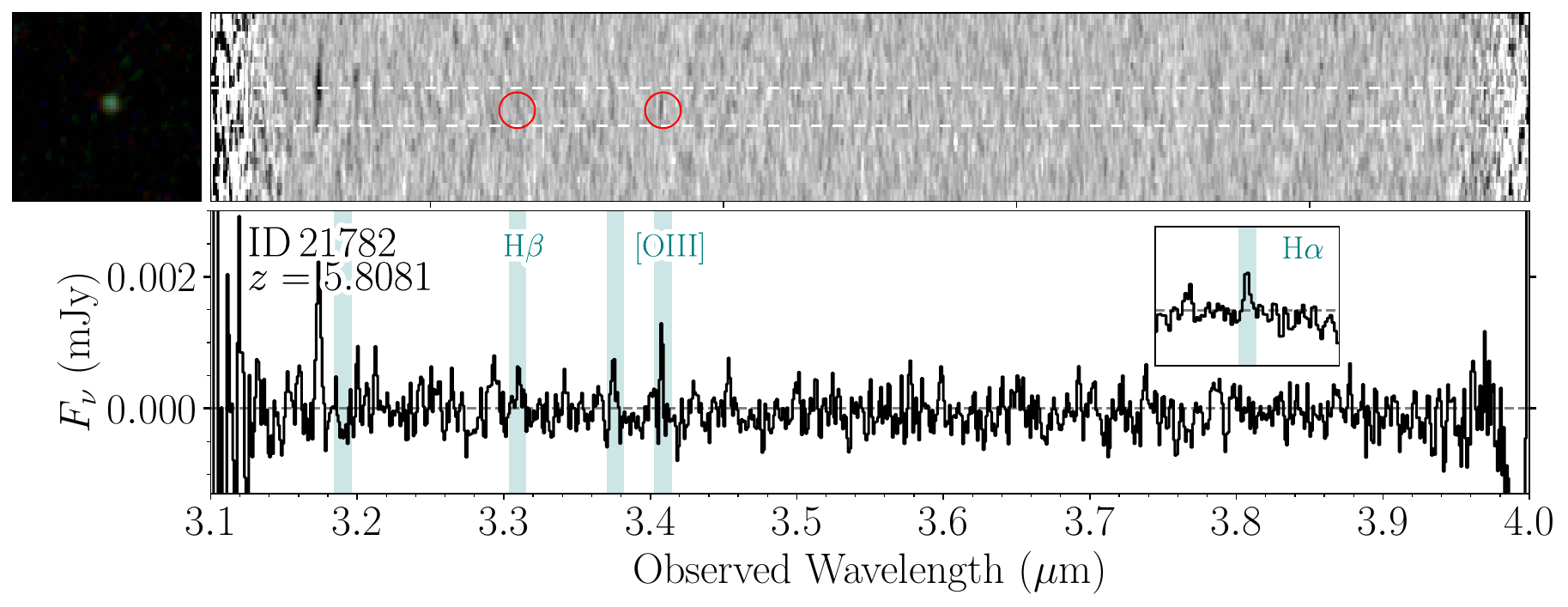}
    \end{minipage}
    \begin{minipage}[b]{0.35\textwidth}
        \includegraphics[width=\textwidth]{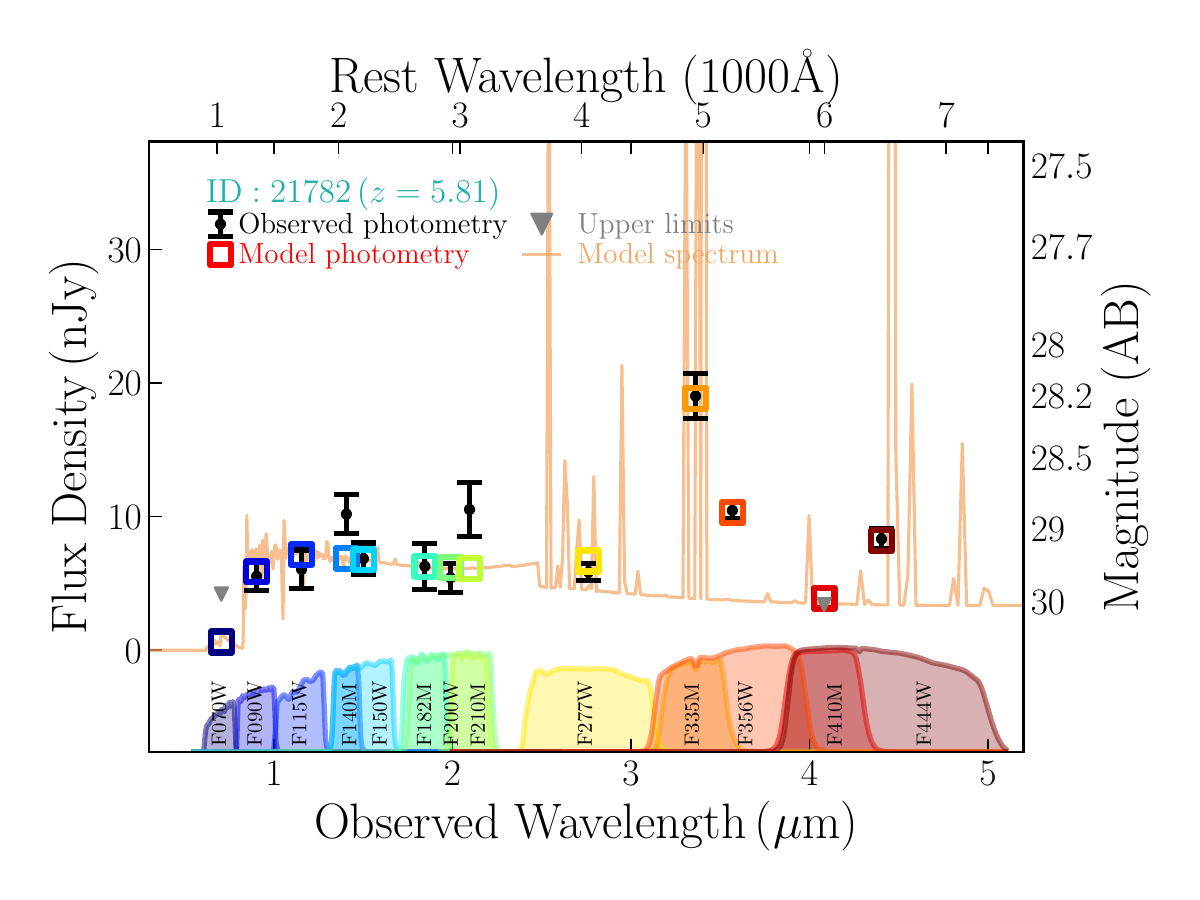}
    \end{minipage}
    \vspace{0.3cm} 
    \centering
    \begin{minipage}[b]{0.63\textwidth}
        \includegraphics[width=\textwidth]{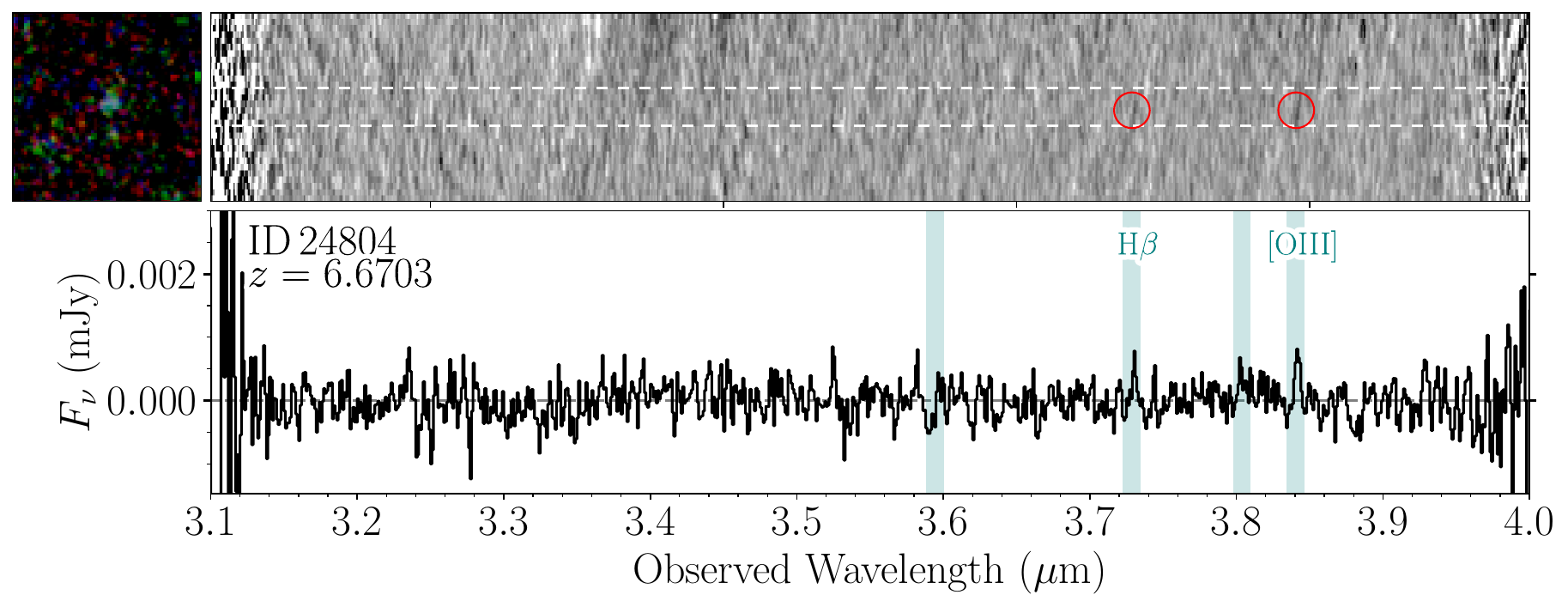}
    \end{minipage}
    \begin{minipage}[b]{0.35\textwidth}
        \includegraphics[width=\textwidth]{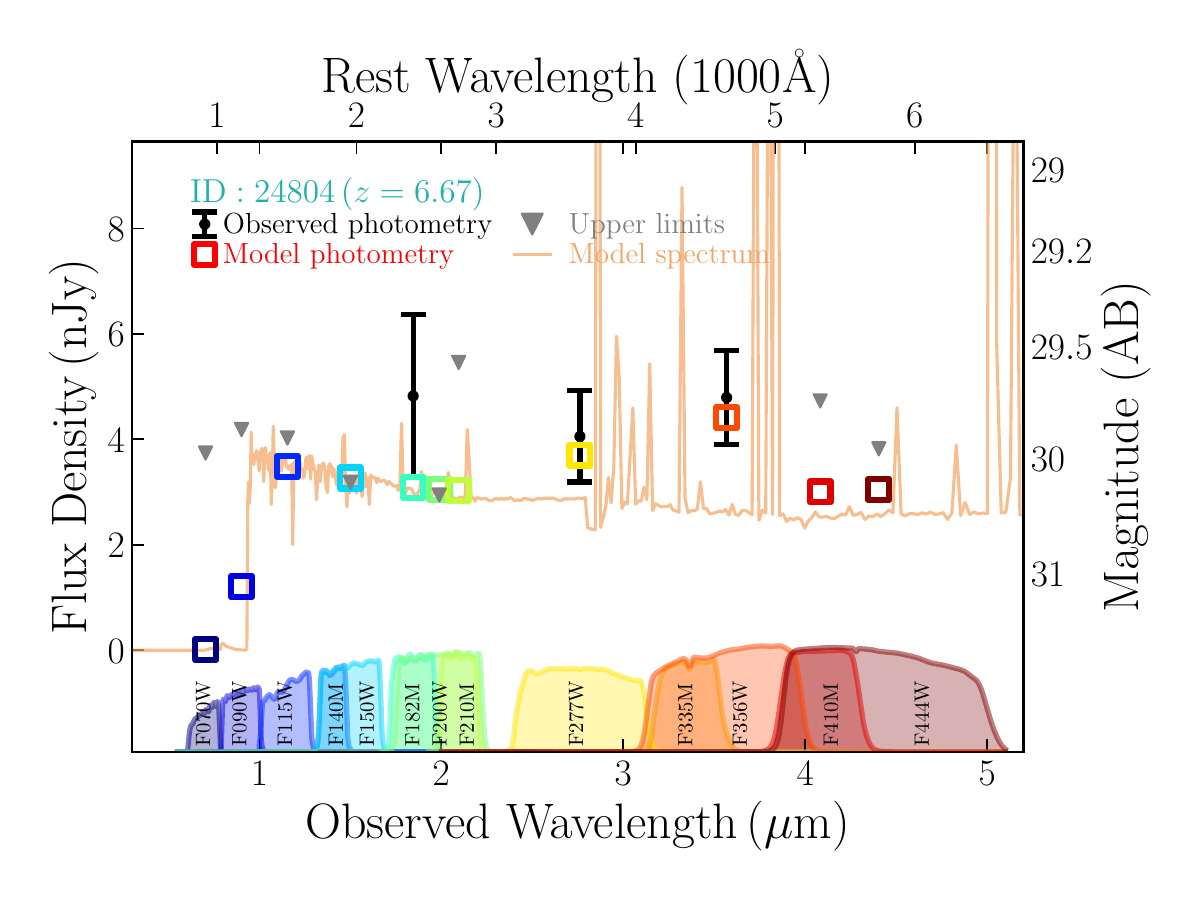}
    \end{minipage}
    \vspace{0.3cm} 
    \centering
    \begin{minipage}[b]{0.63\textwidth}
        \includegraphics[width=\textwidth]{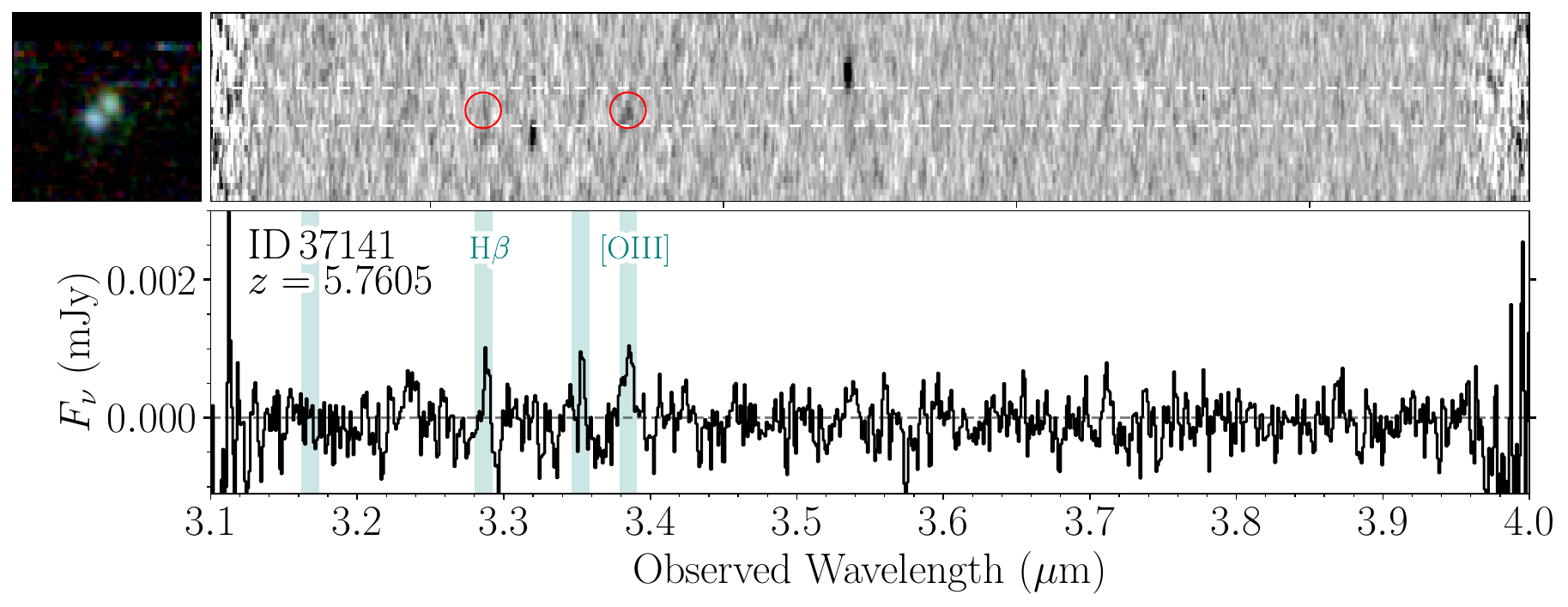}
    \end{minipage}
    \begin{minipage}[b]{0.35\textwidth}
        \includegraphics[width=\textwidth]{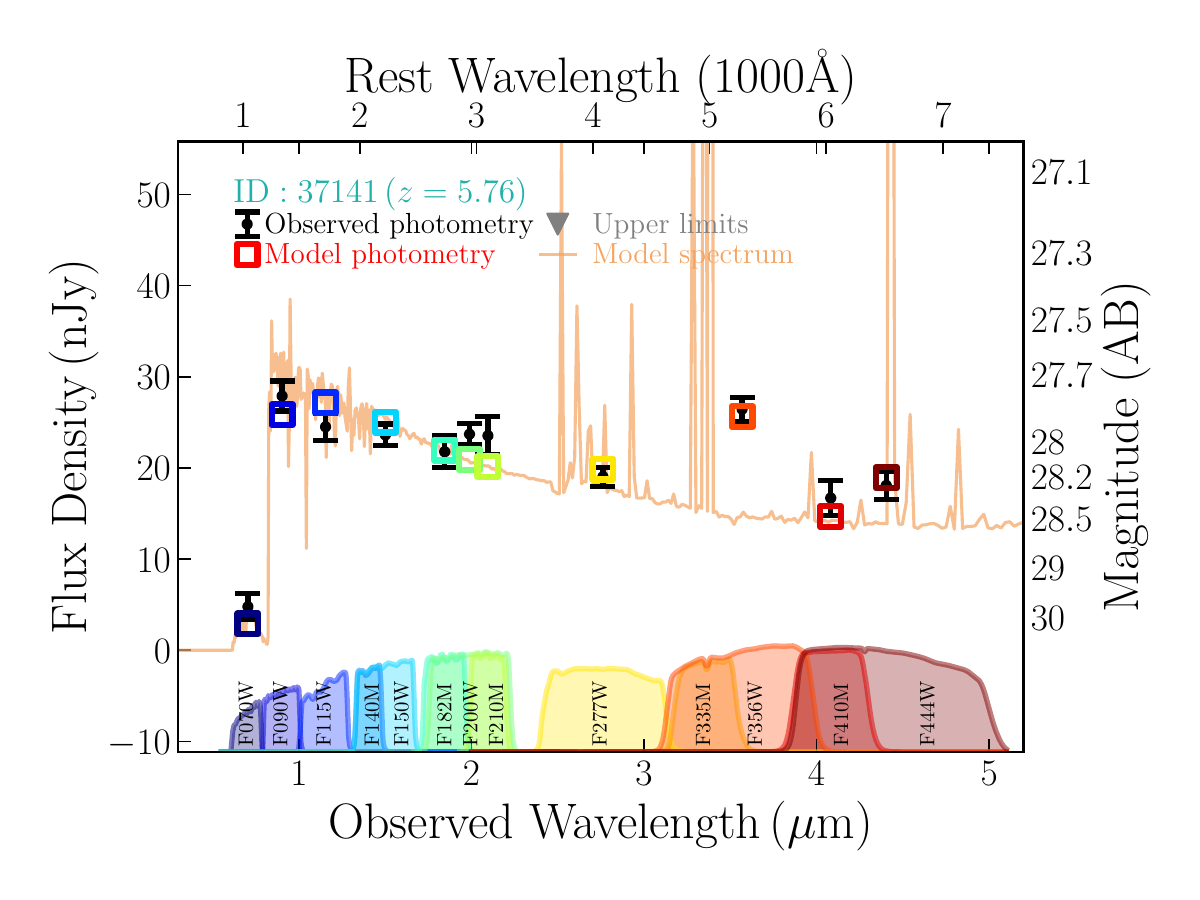}
    \end{minipage}
    \vspace{0.3cm} 
\caption{(Continued from Figure \ref{fig:spec1}.)
\label{fig:spec2}}
\end{figure*}

\section{Results and Discussion}
\label{Sec:discussion}

\subsection{Metallicity Floor?}
\label{Sec:floor}

In the local universe, observations have reached a lower bound in the metallicity distribution, often referred to as the ``metallicity floor'', at $Z\lesssim1\%Z\,_{\odot}$ or $12+{\rm log(O/H)}\lesssim6.69$ \citep[e.g.,][]{Prochaska2003,Thuan2005,Wise2012,Kirby2013}.
With the advent of JWST, it is now possible to search for metal-free or extremely metal-poor galaxies at high redshift, where young, low-mass galaxies are expected to be more prevalent. 
However, even at $z > 5$, nearly all galaxies observed so far continue to exhibit metallicities above this floor \citep[$\gtrsim1-2\%\,Z_{\odot}$; e.g.,][]{Nakajima2023,Curti2024,Sanders2024}.

In this study, we report the discovery of seven EMPG candidates with ${\rm R3}\lesssim3$, corresponding to $Z\lesssim2\%Z\,_{\odot}$ or $12+{\rm log(O/H)}\lesssim7.0$.
Figure \ref{fig:R3} illustrates the strong-line calibration of the R3 ratio as a function of metallicity.
We include various strong-line relations from the literature \citep{Curti2020,Nakajima2022,Hirschmann2023,Sanders2021}, along with our EMPG candidates, calibrated using the \citet{Sanders2024} relation, and other well-studied systems with direct metallicity measurements \citep{Hsiao2024b,Curti2024b,Fujimoto2024,Mowla2024,Alvarez2024,Cullen2025}.
{We also show LAP1-B \citep{Vanzella2023,Nakajima2025}, which {is one} of the most metal-poor systems known at $z>5$}.
It is important to note the considerable scatter in the strong-line relation.
For example, EXCELS-63107 and the Firefly Sparkle appear even more metal-poor than expected given their high R3 values, assuming any of the listed relations.
This scatter suggests that our EMPG candidates may be more metal-poor, or possibly more metal-rich, than inferred, although their true metallicities await confirmation from future direct measurements.
Additionally, high-redshift R3–metallicity calibrations (e.g., \citealt{Hirschmann2023}; \citealt{Sanders2024}) tend to yield lower metallicities at a given R3 compared to low-redshift calibrations (e.g., \citealt{Curti2020}; \citealt{Nakajima2022}). 
\citet{Nakajima2022} further divide their sample by H$\beta$ equivalent width (EW), showing that strong emitters tend to be more metal-poor at a given R3.
{Six of the \texttt{SAPPHIRES} EMPG candidates show high EW(H$\beta)>200\,{\rm \AA}$}.
Therefore, we consider it more appropriate to adopt the high-redshift strong-line calibrations.

To date, only a handful of galaxies {($\#\sim20$)} at $z > 5$ have been reported with ${\rm R3}\lesssim3$ {spectroscopically} {\citep[or $Z\lesssim2\%\,Z_{\odot}$;][]{Vanzella2023,Mowla2024,Chemerynska2024,Vanzella2024,Cullen2025,Willott2025,Fujimoto2025b,Maiolino2025,Nakajima2025,Morishita2025,Korber2026,Asada2026,Caputi2026,Isobe2026,Hsiao2026b}}. 
Our sample {increases the number by 1/3} of such known systems, enabling the first statistical study of EMPGs at high redshift.
{Three of our EMPG candidates, \texttt{SAPPHIRES}-EDR-7695, -9770 and -21785}, exhibit metallicities of $Z\lesssim1\%Z\,_{\odot}$ or $12+{\rm log(O/H)}\lesssim6.69$, assuming the strong-line calibration from \citealt{Sanders2024}).
These values are likely below the metallicity floor observed both in the local universe ($1\%\,Z_{\odot}$) and at high-redshift ($2\%\,Z_{\odot}$), demonstrating the power of deep grism spectroscopic observations.
The remaining EMPG candidates show metallicities comparable to those reported in \citet{Chemerynska2024}, which also adopted the strong-line diagnostics from \citet{Sanders2024}.

In this work, we report {three} additional EMPG candidates at $Z<1\%\,Z_{\odot}$, offering new probes below such a metallicity floor. 
However, we also note a few caveats with our measurements of these sources. 
\texttt{SAPPHIRES}-EDR-24804 has a metallicity of $12+{\rm log(O/H)}=6.77^{+0.13}_{-0.14}$.
This object is also the faintest among our candidates, with only two bands detected confidently, suggesting a low-mass nature (see also \S \ref{Sec:MZ}).
While SED fitting yields a stellar mass of approximately $\sim10^{7.5}\,M_{\odot}$, the limited photometric detections likely leads to an overestimate, and measurements from larger apertures suggest a mass of $<10^{7}\,M_{\odot}$, leaving its physical properties relatively unconstrained.

It is also important to note that LAP-1 was not detected in NIRCam imaging, but was identified purely through emission lines, including Ly$\alpha$, H$\beta$, and \OIIIw\ \citep{Vanzella2023,Nakajima2025}.
This type of search is well-suited for deep grism spectroscopy, but poses significant challenges, as identifying galaxies solely from their line emission requires additional effort and often lacks robust photometric confirmation.
For this reason, in this study, we focus exclusively on the spectra from NIRCam sources that are bright enough to yield meaningful spectral line detections with the grism spectroscopy.
This ensures more reliable estimates of physical properties through SED fitting and enables more secure redshift identification.

We also note that the derived metallicities are based on strong-line empirical calibrations, which can differ substantially depending on the adopted relation. 
Under the most conservative calibration from \citet{Nakajima2022}, which is not specifically calibrated for high-redshift galaxies, none of our EMPG candidates fall below $12+{\rm log(O/H)}\lesssim6.69$.
This highlights the systematic uncertainty associated with applying the calibrations in the local Universe to the early Universe.
Additionally, increased electron densities and compact galaxies at high-redshift \citep[e.g.,][]{Isobe2023,Stiavelli2023,Abdurrouf2024}, could lead to collisional deexcitation of \OIIIw\ and thus decrease the R3 ratio \citep[e.g.,][]{Baskin2005,Katz2023,Zavala2025,Hsiao2026b}.
{Indeed, a recent study has discovered EMPG impostors due to high-electron densities at $z\sim5-6$ \citep{Hsiao2026b}.}
While we do not explore this scenario in detail here, we acknowledge it as a possible factor influencing the observed line ratios, {and more data, such as detections of \OIIIwb\ is essential to measure the metallicity robustly.}

Although many of the most metal-poor galaxies
rely on strong-line empirical calibrations, there are ongoing discoveries of high-redshift EMPGs with the gold-standard, direct $T_{e}$-based metallicity measurements.
\citet{Cullen2025} reported one of the most metal-poor systems to date, EXCELS-63107 ($z=8.271$), with a direct metallicity of $12+{\rm log(O/H)}=6.89^{+0.26}_{-0.21}$ ($\sim1.6\%\,Z_{\odot}$) and an ${\rm R3}=3.61\pm0.31$.
Notably, this R3 value is even higher than expected from high-redshift strong-line calibrations, for example, R3 $= 3.61$ corresponds to $12+{\rm log(O/H)}=7.14$ using our fiducial calibration from \citet{Sanders2024}.
This suggests that our adopted strong-line calibration is not overly aggressive in assigning low metallicities.
These findings underscore the importance of obtaining direct metallicity measurements for EMPGs at high redshift in future surveys (see \S\ref{Sec:future}), to better calibrate strong-line diagnostics and reduce systematic uncertainties.

It is also important to address why such EMPGs with low R3 ratios have not been detected in JWST grism observations from Cycle 1 to Cycle 3, despite plenty of \OIIIw\ emitters discovered in these programs.
For example, compared to other grism surveys such as FRESCO \citep{Oesch2023}, the \texttt{SAPPHIRES}-EDR dataset provides significantly deeper observations.
In F444W, \texttt{SAPPHIRES}-EDR includes 25,510 seconds of exposure in each grism direction (R and C), compared to 7,043 seconds in FRESCO, translating to more than twice the sensitivity in F444W.
As a result, it is not surprising that the 137 \OIIIw\ emitters identified in FRESCO \citep[][]{Meyer2024} do not reach such low R3 ratios.
In fact, even within \texttt{SAPPHIRES}-EDR, we do not identify any EMPGs at $z > 7$ in F444W. 
In the F356W band, the \texttt{SAPPHIRES}-EDR reaches a $5\sigma$ unresolved emission-line sensitivity of $0.55\times10^{-18}\,{\rm erg\,s^{-1}\,cm^{-2}}$ at 3.8\,\micron\ (single dispersion direction; \citealt{Sun2025}), which is deeper than that of Cycle-1 EIGER ($10^{-18}\,{\rm erg\,s^{-1}\,cm^{-2}}$ at $5\sigma$; \citealt{Matthee2023}).
Therefore, we conclude that the depth of \texttt{SAPPHIRES}-EDR is essential for the detections of EMPGs at $z\sim6$.

On the other hand, NIRSpec observations reach significantly deeper sensitivities and, in principle, should be capable of detecting fainter and more metal-poor galaxies.
JADES represents one of the largest and deepest NIRSpec programs to date, having spectroscopically confirmed $\sim$4,000 galaxies \citep{DEugenio2025}.
We examined the JADES Data Release 3 and identified approximately six galaxies with R3 ratios {of 1 $\lesssim$ R3 $\lesssim$ 2,} similar to those of our EMPG candidates, out of a sample of $\sim$300 \OIIIw\ emitters at $z>5.5$.
This suggests that EMPGs can indeed be detected through deep NIRSpec spectroscopy using traditional follow-up methods, but that such sources may have been overlooked ({see also \citet{Hsiao2026b} where they discovered galaxies with four galaxies with R3 $\lesssim$ 5 with NIRSpec}). 
A detailed search or targeted analysis with these dataset is yet to be conducted, which is beyond the scope of this paper.
Nonetheless, this comparison highlights the strength of \texttt{SAPPHIRES}-EDR, as deep NIRCam WFSS exposures over $\sim$16\,arcmin$^2$ yield a comparable number of EMPG candidates to what is found in spectroscopically targeted sample in JADES field ($\gtrsim100$\,arcmin$^2$).


\subsection{Mass-Metallicity relation}
\label{Sec:MZ}

\begin{figure}
\centering
\includegraphics[width=\columnwidth]{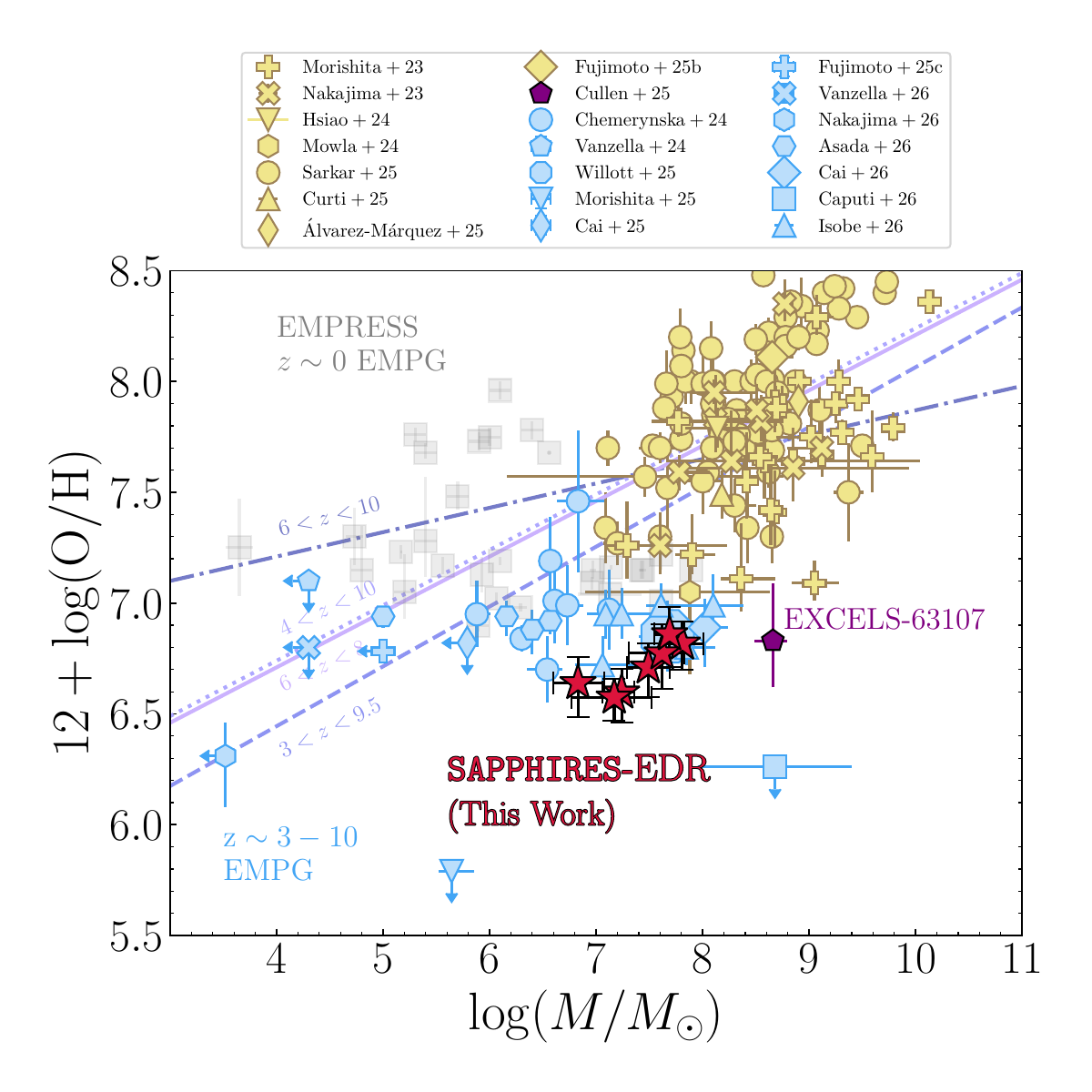}
\caption{The mass-metallicity relation.
The crimson stars represent our EMPG candidates.
We also highlight high redshift galaxies with direct metallicities, as the same markers in Figure \ref{fig:R3} {but in yellow here: MACS0647$-$JD \citep[the upside-down triangle;][]{Hsiao2024b}, the Firefly Sparkle \citep[the hexagon;][]{Mowla2024}, JADES-GS-z9-0 \citep[the triangle;][]{Curti2024b}, GN-z11 \citep[the thin diamond;][]{Alvarez2024}, the Cosmic Grapes \citep[the diamond;][]{Fujimoto2024}, with EXCELS-63107 \citep[the purple pentagon;][]{Cullen2025}, and other EMPGs at $z\sim3-10$ shown in blue: \citet{Chemerynska2024}: the circles, T2c \citep[the pentagon;][]{Vanzella2024}, CANUCS-A370-z8-LAE \citep[the octagon;][]{Willott2025}, AMORE-6 \citep[the upside-down triangle;][]{Morishita2025}, MPG-CR3 \citep[the thin diamond;][]{Cai2025}, GLIMPSE-16043 \citep[the plus;][]{Fujimoto2025,Fujimoto2025b}, LAP-2 \citep[the cross mark;][]{Vanzella2026}, LAP-1 \citep[the hexagon;][]{Vanzella2023,Nakajima2025}, \citet{Asada2026}: the flat-top hexagon, CAPERS-39810 \citep[the diamond;][]{Cai2026}, Pseudo-LRD-NOM \citep[the square;][]{Caputi2026} and \citet{Isobe2026}: the triangles.
Additionally, EMPGs in the local universe literature are also shown: $z\sim0$ \citep[the gray squares;][]{Isobe2022}, as well as other high-redshift galaxies observed with JWST: $4<z<10$ \citep[the blue dotted-line and yellow cross marks;][]{Nakajima2022}, $4<z<10$ \citep[the blue dashed-line and yellow pluses;][]{Morishita2024}, $6<z<10$ \citep[the blue dotted-dashed-line;][]{Curti2024}, $4<z<10$ \citep[the yellow circles;][]{Sarkar2025}{, and $6<z<8$ \citep[the blue solid-line;][]{Hsiao2026a}}.}
}
\label{fig:MZ}
\end{figure}

The stellar mass versus gas-phase metallicity relation, commonly referred to as the mass–metallicity relation, provides one of the most fundamental insights into the galaxy evolution, the baryon cycle, and the regulation of star formation and feedback processes in galaxies \citep[e.g.,][]{Finlator2008,Dave2011,Lilly2013,Kewley2019}.
A tight correlation between stellar mass and gas-phase oxygen abundance of star-forming galaxies in the local Universe has been identified \citep[e.g.,][]{Lequeux1979,Tremonti2004,Lee2006}.
This relation evolves up to $z\sim3$, with lower metallicities at a given stellar mass \citep[e.g.,][]{Erb2006a,Erb2006b,Maiolino2008,Zahid2011,Zahid2013,Andrews2013,LiM2023}. 

Figure \ref{fig:MZ} presents the mass–metallicity relation for our sample, alongside measurements from the literature, including both direct metallicity estimates and those inferred from R3 ratios.
We compare our EMPG candidates with galaxies at similar redshifts observed with JWST, including sources from \citet{Nakajima2023,Curti2024,Morishita2024}, as well as EMPGs at $6 < z < 8$ from \citet{Chemerynska2024},  and local EMPGs from \citet{Isobe2022}.
Compared to local EMPGs, it is evident that high-redshift EMPGs are more metal-poor at a fixed stellar mass, consistent with trends observed in other high-redshift EMPG studies.
This evolution in the mass–metallicity relation from $z \sim 0$ to $z \sim 3$ has been well explored \citep[e.g.,][]{Sanders2021,LiM2023,HeX2024}, and our results further support the existence of such evolution at the low-metallicity end of the relation.

Broadly speaking, our EMPG candidates appear more metal-poor (as intended by our selection) and less massive than typical galaxies at similar redshifts that have been studied with the JWST spectroscopy. 
However, they exhibit a slight offset toward higher stellar mass than might be expected for such low metallicities, particularly when compared to EMPGs at $6 < z < 8$ \citep{Chemerynska2024} and to extrapolations from the high-mass end of the mass–metallicity relation \citep[e.g.,][]{Nakajima2023,Curti2024,Sarkar2025}.
This trend is perhaps not surprising.
Although our sample was not pre-selected based on colors (e.g., for targeted NIRSpec follow-up), the \texttt{SAPPHIRES} photometry does not reach the same depth as some of the deepest JWST programs (e.g., JADES).
As a result, while we aim to identify extremely metal-poor galaxies, our selection remains biased toward brighter and more massive EMPGs \citep[Malmquist bias;][]{Malmquist1922}. Additionally, intrinsic scatter in the mass–metallicity relation contributes to this spread.
It is also unsurprising that \citet{Chemerynska2024} reported a mass–metallicity relation broadly consistent with expectations from the high-mass end, as their primary objective was to probe the low-mass end of the relation.
Furthermore, their sample was drawn from the UNCOVER survey in the Abell 2744 field, where the high-redshift galaxies are gravitationally magnified, enabling the detection of intrinsically fainter and lower-mass galaxies.
  
We also highlight the presence of one extremely metal-poor system, identified through the gold-standard, direct $T_{e}$-based metallicity measurements, that exhibit similar metallicities and stellar masses to our EMPG candidates. 
EXCELS-63107 \citep{Cullen2025}, for instance, has an even higher stellar mass of ${\rm log}(M/M_{\odot})=8.66^{+0.13}_{-0.17}$, compared to galaxies at similar metallicities. This suggests that metal-poor, yet moderately massive galaxies can exist at high redshift.
These findings underscore that the mass–metallicity relation may not be strictly monotonic, particularly at the low-metallicity end {(see also \citealt{Hsiao2026a})}.
That said, determining whether EMPGs follow distinct strong-line relations will require future, deeper, and wider spectroscopic surveys (see \S\ref{Sec:future}).

\subsection{Star-forming main sequence}
\label{Sec:SFR}

The so-called ``star-forming main sequence'' (SFMS) describes the relationship between the SFR and stellar mass $M_{*}$ of star-forming galaxies across a wide range of redshifts.
The main sequence depicts that most star-forming galaxies evolve in a relatively steady, secular mode of star formation, rather than through stochastic or bursty events. 
The slope and normalization of the SFMS evolve with redshift, with galaxies at higher redshifts exhibiting higher SFRs at fixed stellar mass.
The SFMS provides a useful benchmark for interpreting galaxy evolution, as deviations above or below the sequence can indicate starburst activity or quenching, respectively.
In this context, the placement of EMPGs on or off the SFMS can offer insights into their star formation modes, gas content, and evolutionary states at early cosmic times.

\begin{figure}
\centering
\includegraphics[width=\columnwidth]{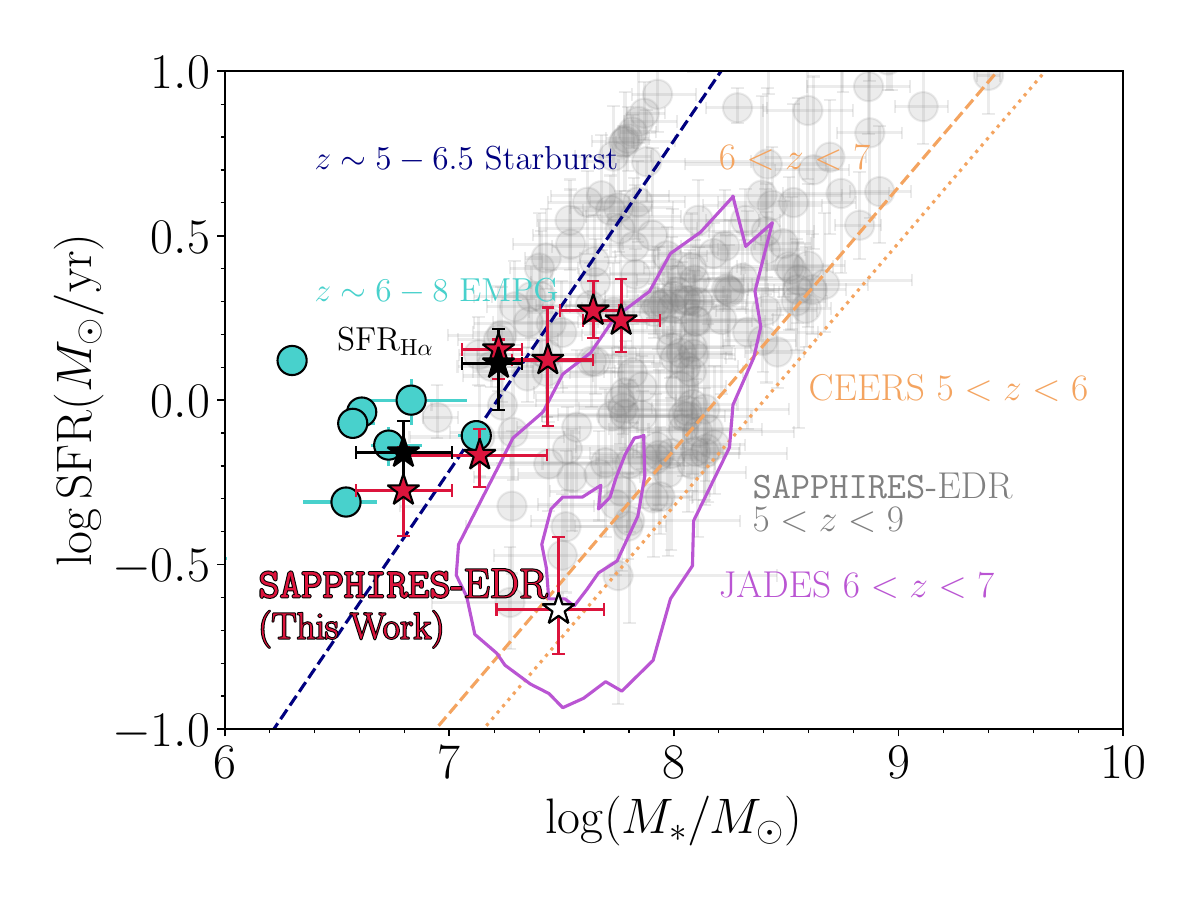}
\caption{Star-formation main sequence (i.e., SFR vs $M_*$) of EMPG candidates.
The crimson stars show the SFR$_{\rm SED}$ while the black stars represent the SFR$_{\rm H\alpha}$.
Other high-z EMPGs are shown in the cyan circles for comparison \citep{Chemerynska2024}.
The orange line demonstrate the linear fits of star-formation main sequence from CEERS data \citep{Cole2025} at $z\sim5-6$ (dotted-line) and $z\sim6-7$ (dashed-line), while purple contours indicate the 85\% and 95\% of star-formation main sequence from JADES data \citep[$z\sim6-7$;][]{Ciesla2024}, as well as the whole population of \texttt{SAPPHIRES}-EDR at $z\sim5-9$ in the gray circles \citep{Sun2025}.
The relation of high-redshift starburst galaxies is also shown in navy dashed-line \citep{Rinaldi2022}.
}
\label{fig:SFRM}
\end{figure}

Our EMPG candidates have stellar masses of $10^{6.8}-10^{7.8}\,M_{\odot}$ and SFRs of $0.2-2\,M_{\odot}/{\rm yr}$, corresponding to specific star formation rates of $\sim5-100\,{\rm Gyr^{-1}}$.
We present the star-formation main sequence in Figure \ref{fig:SFRM}.
We note that our SFRs are derived in two ways: 1.) from SED fitting, averaged within $10\,{\rm Myr}$ before the epoch of observation, and 2.) from H$\alpha$ if available (see next paragraph).
Previous studies used different timescales and methods to estimate the mass and SFR.
We compare our EMPG candidates to galaxies at similar redshifts ($z\sim 5-7$ with JWST observations, including JADES, \citealt{Ciesla2024}; and CEERS, \citealt{Cole2025}).
Most of our EMPG candidates, except 24804, display $\sim0.5\,{\rm dex}$ higher SFR at a given stellar mass, suggestive of more bursty SFR, compared with most of the star-forming galaxies, and of course with quenched galaxies \citep[e.g.,][]{Looser2024}.
{Such bursty nature can be also seen in the small $\tau\sim5\,{\rm Myr}$, presented in Table \ref{tab:SED}.}
Although as a grism survey, \texttt{SAPPHIRES}, may already be biased to samples with high sSFR (i.e., strong line emitters), {our additional requirement of S/N $\geq 3$ for the emission lines used in the metallicity determination may further bias the EMPG sample toward galaxies with stronger nebular emission and hence higher SFRs.
Consequently, galaxies with similarly low metallicities but weaker star formation may fail the line-detection criterion and be underrepresented, particularly in the lower-left region of Figure~\ref{fig:SFRM}.
Nevertheless,} they are overall consistent (possibly with a flatter slope) compared with other bursty star-forming galaxies at similar redshift \citep{Rinaldi2022}.
The similar bursty nature was also seen in other high-redshift EMPGs \citep{Chemerynska2024}, albeit they primarily adopted the H$\alpha$ SFR, and are even more bursty, than \citet{Rinaldi2022} for example, and also in extreme emission line galaxies \citep{Boyett2024}.
It suggests that there are pristine gas inflows that dilute the gas phase metallicity and increase the SFR simultaneously \citep[][]{Curti2024}.
Unsurprisingly, most of the low-mass ($M<10^{7}\,M_{\odot}$) galaxies at $5<z<7$ in \citet{Rinaldi2025} are extreme starbursts, which is possibly caused by a selection effect, as low-mass galaxies with strong emission lines are easier to capture, as pointed out by \citet{Chemerynska2024}.

We also estimate SFR from the H$\alpha$ emission when detected.
We first compute the intrinsic H$\alpha$ luminosity, and then convert it to SFR using a calibration factor of 3.2 $\times$ \tentotheminus{42} ${\rm M_{\odot}/yr/(erg/s)}$, appropriate for high-redshift, metal-poor galaxies \citep{Reddy2018} for a \citet{Chabrier2003} IMF.
The derived SFR$_{\rm H\alpha}$ values are listed in Table \ref{tab:SED}.
{Both} EMPG candidates with H$\alpha$ detections exhibit SFR$_{\rm H\alpha}$ values that are {consistent with} the corresponding SFR$_{\rm SED}$ , even though SFR$_{\rm SED}$ is averaged over the last $10\,{\rm Myr}$ prior to observation.
As shown in Figure \ref{fig:SFRM}, these sources appear even more bursty than the starburst galaxies studied in \citet{Rinaldi2022}.
This behavior is consistent with trends found in \citet{Chemerynska2024}.
Together and in general, this might indicate that, in order to efficiently select EMPGs at high redshift, bursty star-forming and low-mass galaxies can be excellent candidates for follow-up spectroscopy.

We also note that all of the EMPG candidates with H$\alpha$ exhibit Balmer decrements (H$\alpha$/H$\beta$) lower than the Case B recombination value of 2.86 \citep{Osterbrock1989}.
Such low ratios have been observed in other galaxies with JWST \citep{Pirzkal2024, SunB2025}, and may either reflect physical conditions \citep{Yanagisawa2024} or simply result from increased noise in the F444W band.
{Several observational and modeling uncertainties may also contribute to this discrepancy.
First, our selection targets faint galaxies, whose marginal H$\beta$ and H$\alpha$ measurements are susceptible to noise fluctuations in either direction. Selection based on H$\beta$ detections may preferentially retain upward fluctuations, lowering the inferred Balmer decrements.
Second, residual relative flux-calibration or extraction systematics between F356W and F444W, which contain H$\beta$ and H$\alpha$, respectively, may affect the measured ratios. Although \citet{Pirzkal2026} report accurate NIRCam grism flux calibration, this does not exclude additional uncertainties in extracted galaxy spectra. As an analogy, measurements of the same emission line in overlapping NIRSpec gratings can differ by factors of 2--4 in the most extreme cases (\citealt{Hsiao2026b}; Taylor et al., in prep.).
Finally, agreement between H$\alpha$- and SED-derived SFRs does not independently validate the H$\alpha$ fluxes, as their correspondence depends on SFH and the priors adopted in SED fitting.
Deeper Cycle 5 observations will provide more robust Balmer-line measurements to help assess these possibilities (see \S\ref{Sec:future}).}
Regardless, if we adopt H$\beta$ assuming no dust attenuation, the inferred specific star formation rates (sSFRs) of these EMPGs would be even higher.

{We also caution that the assumed delayed-$\tau$ SFH model introduces unavoidable uncertainties in the derived stellar masses, SFRs, and ages.
Given the implied short-lived burst SFH, it is possible that the SED fitting captures only a recent star formation episode, while an older underlying stellar population is outshined by the younger, brighter stars \citep[e.g.,][]{Leja2019a,Leja2019b,Gimenez2023,Gimenez2024,Bradac2024}.
Therefore, our stellar mass might potentially be underestimated, and would bring these EMPGs closer to the bulk of the star-forming main sequence in Figure \ref{fig:SFRM}.}

{
We measure the rest-frame UV ($\sim2600–2900\,{\rm \AA}$) sizes on the F200W mosaic by forward-modeling each candidate with the empirical PSF (built from stars in the same mosaics) convolved with a circular 2D Gaussian profile, with uncertainties from Monte-Carlo noise realizations.
Two candidates (7695 and 37141) are marginally resolved with intrinsic half-light radii $r_e\sim170\,{\rm pc}$; four are unresolved with 2$\sigma$ upper limits of $r_e\lesssim105–280\,{\rm pc}$; 24804 is too faint for a meaningful constraint.
The implied SFR surface densities are $\Sigma_{\rm SFR} = 8–11\,M_\odot\,{\rm yr^{-1}\, kpc^{-2}}$ for the resolved candidates and $\gtrsim1–10\,M_\odot\, {\rm yr^{-1}\, kpc^{-2}}$ for the rest. The measured sizes and SFR surface densities imply compact and dense star-forming regions that are comparable to other star-forming systems at similar redshifts \citep[e.g.,][]{Ono2025}.}

\subsection{Age, Dust, and Color}
\label{Sec:properties}

Most of our candidates are estimated, from SED fitting, to be relatively young ($\lesssim25\,{\rm Myr}$), with the exception of SAPPHIRES-EDR-24804 because of the limited data quality.
These age estimates are consistent with the expectations from previous studies of local EMPGs, which typically show ages of $\lesssim50\,{\rm Myr}$ \citep[e.g.,][]{Isobe2022}.
At this timescale, neither metals nor stellar mass have had sufficient time to accumulate significantly.
Similarly young stellar ages have been observed in other high-redshift EMPGs including CANUCS-A370-z8-LAE ($<20\,{\rm Myr}$; \citealt{Willott2025}).
{
Simulations predicted that the lowest-mass dwarf galaxies in the early Universe, which undergo rapid chemical enrichment and initially form from extremely metal-poor gas, have very young stellar populations and short star formation histories \citep{Salvaterra2011,Gelli2020}.} 

We also estimate the rest-frame UV colors, characterized by the UV continuum slope ($\beta$, $f_{\lambda}\propto\lambda^{\beta}$), for our EMPG candidates using the {F115W, F150W, and F200W} photometry {(corresponding to the rest-frame wavelength of $\sim$1700$\,\rm{\AA}$--2500$\,\rm{\AA}$)}. 
The derived $\beta$ values for each candidate are presented in Table \ref{tab:SED}.
We do not attempt to estimate the UV slope for \texttt{SAPPHIRES}-EDR-24804, as it is only detected in two bands.
The remaining EMPG candidates exhibit blue UV slopes, ranging from {$\beta\sim-1.7$ to $-2.8$}, which are bluer than the typical median value of $\beta=-2.0$ reported for galaxies at $7<z<11$ \citep{Topping2022}, and comparable to galaxies at even higher redshifts $9<z<12$ \citep{Cullen2024}.
However, the UV slopes of our EMPG candidates are not as extreme as those of the bluest galaxies known at high redshift \citep[e.g.,][]{Cullen2023,Yanagisawa2024b}.
Notably, none of our sources exhibit {$\beta\lesssim-2.8$}, suggesting that there is no strong evidence that their UV emission is dominated by ultra-young, dust-free stellar populations with high Lyman continuum escape fractions, which has been proposed for some of the most extreme high-redshift systems \citep{Cullen2023}.

Compared to other EMPGs at high redshifts, EXCELS-63107 exhibits an exceptionally blue UV slope of $\beta=-3.3\pm0.3$ \citep{Cullen2025}.
In contrast, CANUCS-A370-z8-LAE shows a significantly redder UV slope of $\beta=-1.80\pm0.15$ \citep{Willott2025}.
While such a red continuum is typically attributed to moderate dust attenuation \citep[e.g.,][]{Hsiao2023a,Fujimoto2024b}, the red UV color of CANUCS-A370-z8-LAE may require more sophisticated modelling to fully explain its peculiar beta slope.

Our EMPG candidates exhibit UV slopes that fall between those of the two systems.
This intermediate range can likely be explained by low but non-zero dust attenuation $A_{V}\sim0.01-0.2\,{\rm mag}$, which result in less extreme UV spectral slopes.
In comparison, local EMPG analogs generally display redder UV slopes, with $-1.9\gtrsim\beta\gtrsim-2.3$ for galaxies with $Z\sim0.1\,Z_{\odot}$ \citep{Isobe2022}.
For those with $Z\sim0.02\,Z_{\odot}$, they have UV slopes of $-2\gtrsim\beta\gtrsim-3$ \citep{Nakajima2022}, which are broadly consistent with the UV slopes measured in our high-redshift EMPG sample.
Therefore, we do not observe a significant trend toward bluer UV slopes when comparing local EMPGs to their high-redshift counterparts.
While the overall galaxy population shows a clear trend toward bluer $\beta$ values at higher redshifts, likely due to lower metallicities, dust content, and younger stellar populations, the EMPG population appears to maintain similar physical conditions across cosmic time.
This consistency is possibly simply because EMPGs are, by selection, dominated by the most metal-poor and youngest stellar populations at any epoch.

In addition to stellar continuum emission, nebular continuum can become significant during extreme starbursts \citep[e.g.,][]{Cameron2024, Katz2024, Saxena2024, Welch2025}, particularly in young, low-mass, and metal-poor galaxies \citep{Izotov2011}, conditions that are consistent with our EMPG candidates.
For most of our EMPGs, the SED fitting results show no significant Balmer jump, suggesting that the nebular continuum contribution may be negligible.
In contrast, \texttt{SAPPHIRES}-EDR-9770 exhibits a strong Balmer jump, along with a relatively {blue} UV slope {($\beta \sim -2.7$)}.
If nebular continuum emission is present, it can redden the UV slope by approximately 0.1–0.5 dex \citep{Schaerer2002, Cullen2024, Katz2024}.

\citet{Fujimoto2025} utilized NIRCam color-color diagrams to select candidate Pop {\sc III} galaxies across $\simeq500\,{\rm arcmin^{2}}$ of JWST fields.
Their approach involved a novel and careful selection of galaxies at redshifts $z\sim6-7$, based on several predicted Pop {\sc iii} features: the absence of \OIIIw, the presence of strong hydrogen recombination lines (H$\alpha$ and H$\beta$), and a pronounced Balmer jump.
They identified one promising candidate, GLIMPSE-16043, that satisfied all selection criteria \citep[see also][]{Zackrisson2011}.
{Although $\sim$40 hours of spectroscopic follow-up under JWST DDT program 9223 (PI: Fujimoto) ruled out a metal-free interpretation due to a stronger Balmer jump than expected \citep{Fujimoto2025b}, the color-color selection remains useful for finding metal-poor galaxies.}

\begin{figure*}
\centering
\includegraphics[width=\textwidth]{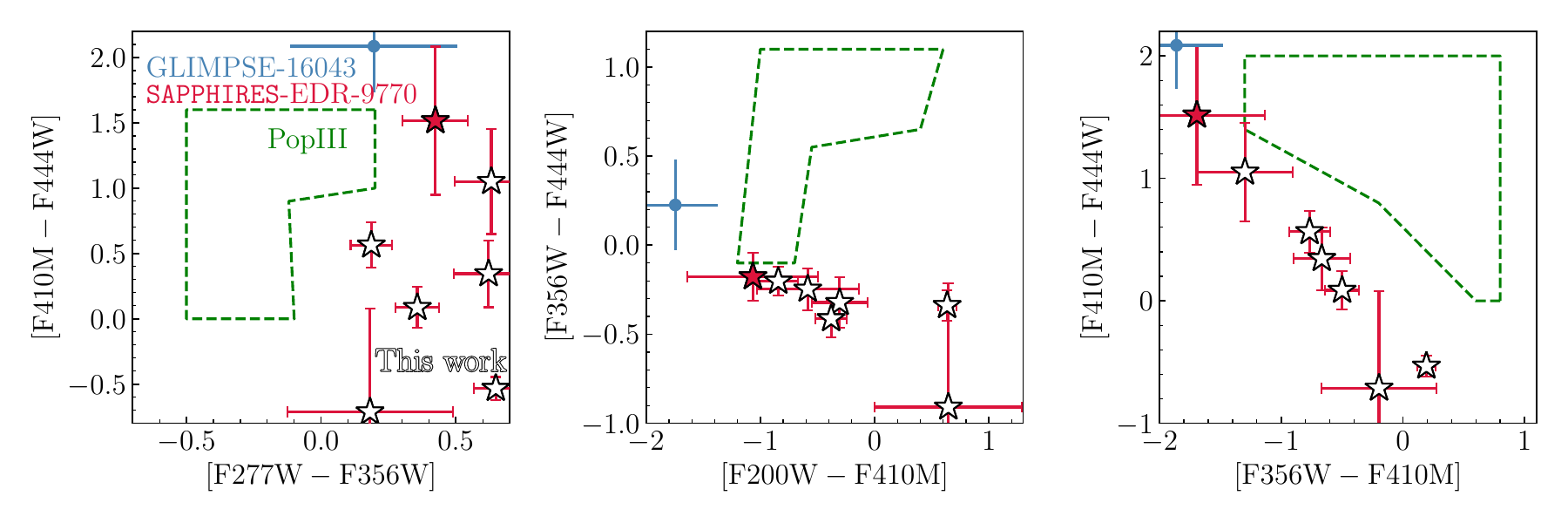}
\caption{Color-color diagrams compared with Pop {\sc iii} models and the Pop {\sc iii} candidate, GLIMPSE-16043 in \citet{Fujimoto2025,Fujimoto2025b}.
Our EMPG candidates are shown in the crimson stars, and the Pop {\sc iii} candidate GLIMPSE-16043 is shown in the blue dots.
The regions where the green dashed-lines enclosed represent the Pop {\sc iii} selection from \citet{Fujimoto2025,Fujimoto2025b}.
}
\label{fig:color}
\end{figure*}

We compare the colors of our EMPG candidates against the color–color selection criteria defined in \citet{Fujimoto2025,Fujimoto2025b}, as shown in Figure \ref{fig:color}.
Weak \OIIIw\ and/or the Balmer jump are reflected in x-axis, while y-axis colors are sensitive to the the strong Balmer line feature (i.e., H$\alpha$).
The majority of our EMPG candidates do not fall within the Pop {\sc iii} selection region, which is expected because these galaxies are partially selected with \OIIIw\ emission, resulting in enhanced flux in the F356W band. 
The presence of \OIIIw\ shifts the colors such that [F277W--F356W] is elevated, while [F356W--F444W] and [F356W--F410M] are suppressed.
Additionally, it is possible that H$\alpha$ is not strong enough in these galaxies to significantly impact the y-axis colors.
Nevertheless, we identify one source, \texttt{SAPPHIRES}-EDR-9770, that shows consistency with all three Pop {\sc iii} color–color selection criteria {in the old \citet{Fujimoto2025} color-color selection, within the uncertainties, but in the latest \citet{Fujimoto2025b} version, \texttt{SAPPHIRES}-EDR-9770 is only consistent with two criteria. 
}
This suggests that our grism spectroscopic selection of EMPG candidates is approaching (but not yet reaching) the regime of Pop {\sc iii} galaxies.

\subsection{Stacked Spectrum}
\label{Sec:stack}

\begin{figure}
\centering
\includegraphics[width=\columnwidth]{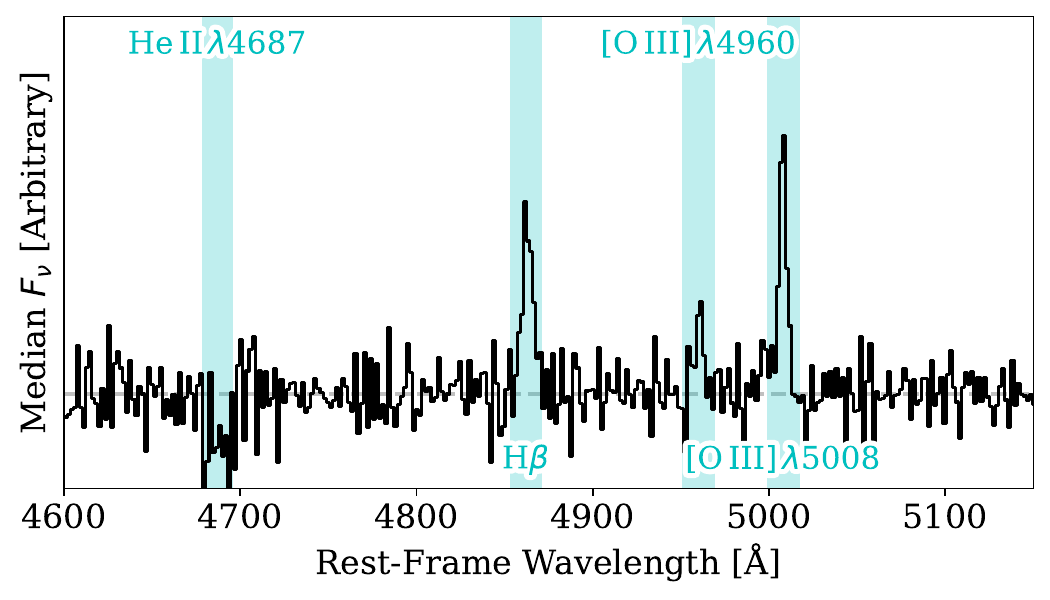}
\caption{Stacked spectrum of seven EMPG candidates at 4600--5150\,\AA.
}
\label{fig:stack}
\end{figure}

We also construct a simple median stacked spectrum of our seven EMPG candidates{, with each spectrum shifted to the rest frame according to the redshift measured from the \OIIIw\ centroid,} to search for faint spectral features that may not be visible in individual spectra. 
Figure \ref{fig:stack} presents the stacked spectrum.
Notably, \OIIIwc\ becomes clearly visible in the stack, even though it is not distinctly detected in several individual spectra.
From the stacked spectrum, we estimate the \OIIIw/\OIIIwc\ flux ratio to be $3.1\pm0.5$, consistent with theoretical expectations of 2.98 \citep{Storey2000}, supporting the interpretation that these sources are indeed \OIIIw\ emitters.
The H$\rm\beta$ line is also detected at high confidence in the stacked spectrum of SNR $\sim12$, and we measure an R3 of $1.3\pm0.1$, consistent with the median of individual R3.

In addition, we highlight the  \HeIIwb\ region in Figure \ref{fig:stack}.
\HeIIwb\ is expected in the spectra of metal-free galaxies or even in extremely metal-poor Pop II galaxies  \citep[e.g.,][]{Oh2001,Nakajima_popIII}.
However, we do not detect \HeIIwb\ in any individual spectrum, nor in the stacked spectrum.
Following the diagnostic framework of \citet{Nakajima_popIII}, where \HeIIwb/H$\beta\sim0.1$ for Pop {\sc III} and \HeIIwb/H$\beta\sim0.01$ for Pop {\sc II} EMPGs, we note that the integrated H$\beta$ in the stacked spectrum has a SNR $\sim12$.
Given this relatively modest SNR, it is not surprising that \HeIIwb\ is not detected, even if some of these galaxies host Pop {\sc III}-like stellar populations.
A possible absorption feature might be simply affected by 21782, as \HeIIwb\ is hitting the edge of the transmission.
Further deeper NIRSpec observations are needed to detect faint lines like \HeIIwb.

\subsection{{Future Observations}}
\label{Sec:future}
Throughout this paper, we adopt strong-line ratios to estimate gas-phase metallicities (see \S\ref{Sec:strongline}).
However, we emphasize that strong-line calibrations at high redshift remain poorly constrained, with only $<100$ galaxies having direct metallicity measurements to date \citep[e.g.,][]{Sanders2024,Laseter2024,Sarkar2025}.
To robustly confirm the metallicity of our candidates, future deep NIRSpec observations will be essential to detect temperature-sensitive auroral lines such as \OIIIwa\ and \OIIIwb, as well as to resolve density-sensitive doublets like \OIIdw\ and \CIIIdw.
{JWST Cycle 5 GO 10631 (PI Hsiao) will use NIRSpec G235H and G395H to further detect \OIIIwa\ and resolve \OIIdw\ in these EMPG candidates.
Future G140H observations will be also crucial to detect \OIIIwb\ and measure their electron densities in the high-ionization zone to confirm whether they are real EMPGs or impostors \citep[see][]{Hsiao2026b}.
}
These observations will enable direct measurements of electron temperature and electron density, allowing for more accurate determinations of gas-phase metallicity.

In parallel, ongoing and future efforts to improve the strong-line calibrations at high redshift will be crucial, particularly for application to larger samples. For example, JWST Cycle 4 program GO-7729 (PI: Roberts-Borsani, G.) is expected to significantly expand the number of direct metallicity measurements to over 70 galaxies at $z>6$ (similar for GO-7935; PIs: Sun, F. \&\ Lin, X.).

Deeper NIRCam and MIRI imaging will also be essential for improving constraints on stellar masses.
For instance, our most metal-poor candidate, SAPPHIRES-EDR-24804, is detected only in F277W and F356W, leading to large uncertainties in the SED-derived physical parameters due to limited wavelength coverage.
{JWST imaging would also enable a broader photometric search for EMPG candidates \citep[e.g.,][]{Fujimoto2025,Trussler2026}.}

Finally, our work highlights the power of deep NIRCam/WFSS parallel observations in identifying extremely metal-poor galaxies. Leveraging one of the deepest JWST grism observations to date, we identify seven EMPG candidates, nearly doubling the number of such sources discovered previously.
Future \texttt{SAPPHIRES} observations, along with potential ultra-deep grism programs, promise to uncover even more extremely metal-poor systems at high redshift and may ultimately transform our understanding of star formation in low-metallicity or Pop {\sc iii} environment.

\section{Conclusions} \label{sec:conclusion}
In this paper, we search for extremely metal-poor
galaxies (EMPGs; $Z\lesssim2\%\,Z_{\odot}$) in the JWST Cycle 3 Treasury Program, the Slitless Areal Pure-Parallel HIgh-Redshift Emission Survey (\texttt{SAPPHIRES}, GO‑6434; PI: Egami et al.).
The ultra-deep grism data from SAPPHIRES Early Data Release (EDR; \citealt{Sun2025}) is suitable for a flux-complete search of EMPG candidates at $z\simeq5-9$.
We fit the \OIIIw\ and H$\beta$, for all 120 \OIIIw\ emitters at $z\simeq5 - 9$, and select EMPG candidates based on low R3=\OIIIw/H$\beta$ ratio.
We summarize the main results below:
\begin{enumerate}
    \item We discover seven EMPG candidates with ${\rm R3}\lesssim3$, corresponding to $Z\lesssim2\%Z\,_{\odot}$ or $12+{\rm log(O/H)}\lesssim7.0$.
    Among them, {three} candidates exhibit even lower metallicities of  $Z\lesssim1\%Z\,_{\odot}$ or $12+{\rm log(O/H)}\lesssim6.69$.
    Their existence breaks the so-called ``metallicity floor'' observed in the local universe ($1\%\,Z_{\odot}$) and at high redshifts ($2\%\,Z_{\odot}$).
    These EMPG candidates appear extremely faint ($\sim28-30\,$AB mag) and low-mass (${\rm log(M_{*}/M_{\odot})}\sim6.8-7.8$), as expected from the mass–metallicity relation. 
    
    \item These EMPG candidates have 
    star formation rates (SFRs) of $0.2-2\,M_{\odot}/{\rm yr}$, corresponding to specific star formation rates of $\sim5-100\,{\rm Gyr^{-1}}$.
    Six of the seven EMPG candidates exhibit SFRs that are $\sim0.5\,{\rm dex}$ higher than typical galaxies at the same stellar mass and redshift, suggesting more bursty star formation, consistent with the behaviours of other high-redshift EMPGs.

    \item Six EMPG candidates are estimated to be relatively young ($\lesssim25\,{\rm Myr}$), exhibiting blue UV slopes ranging from {$\beta\sim-1.7$ to $-2.8$},  and low dust attenuation ($A_{V}\sim0.01-0.2\,{\rm mag}$).
    These estimates are generally consistent with those of local and high-redshift EMPGs, as well as with the predictions from certain cosmological simulations.
    We also compare the NIRCam colors of our EMPG candidates using color–color diagrams designed to identify Pop {\sc iii} galaxy candidates.
    Six of the seven EMPG candidates fall outside the Pop {\sc iii} selection region, likely because their SEDs still exhibit detectable \OIIIw\ emission.
    However, one candidate, \texttt{SAPPHIRES}-EDR-9770, shows comparable colors as those of Pop {\sc iii} predictions within uncertainty.
    
    \item We also construct a stacked spectrum of our seven EMPG candidates, in which \OIIIwc\ becomes clearly visible. 
    From the stacked spectrum, we estimate the  \OIIIw/\OIIIwc\ flux ratio to be $3.1\pm0.5$, consistent with theoretical expectations. 
    We also measure the R3 ratio of $1.3\pm0.1$.
    This further supports the interpretation that these candidates are genuine \OIIIw\ emitters.
    The \HeIIwb\ is not detected even in the stacked spectrum,  
    which is expected given the current depth of the observations.
    
\end{enumerate}

Our results demonstrate the effectiveness and power of NIRCam/WFSS in identifying metal-poor galaxies without requiring pre-selection like those for slit spectroscopy.
Future ultra-deep NIRCam/WFSS observations (including SAPPHIRES and GTO-4540, \citealt{Eisenstein2023b}) hold the promise of uncovering even more extremely metal-poor systems at high redshifts and potentially transforming our understanding of Pop {\sc iii} star formation.
In addition, future JWST/NIRSpec follow-up observations will be essential for obtaining gold-standard, direct metallicity measurements {(e.g., GO 10631)}.
These observations will not only confirm the metallicities of the EMPG candidates presented here, but also help improve the strong-line diagnostics at high redshifts for application in large spectroscopic surveys.

\section{Acknowledgments}
{We appreciate the anonymous referee for their constructive comments.} 
We thank Prof. Steven Finkelstein, Prof. Harley Katz, Dr. Rohan Naidu, Dr. Seiji Fujimoto, Dr. James Trussler, and Dr. Tobias Looser for helpful discussions.
This work is based on observations made with the NASA/ESA/CSA James Webb Space Telescope. The data were obtained from the Mikulski Archive for Space Telescopes at the Space Telescope Science Institute, which is operated by the Association of Universities for Research in Astronomy, Inc., under NASA contract NAS 5-03127 for JWST. These observations are associated with program \#6434.
Support for program \#6434 was provided by NASA through a grant from the Space Telescope Science Institute, which is operated by the Association of Universities for Research in Astronomy, Inc., under NASA contract NAS 5-03127.
We are grateful and indebted to the 20,000 people who worked to make JWST an incredible discovery machine.
TH is funded by grants for JWST-GO-4246 provided by STScI under NASA contract NAS5-03127.
TH appreciates the support from the Government scholarship to study abroad (Taiwan).
Y.F.\ is supported by JSPS KAKENHI Grant Numbers JP22K21349 and JP23K13149.
F.S., J.M.H., E.E., D.J.E., C.N.A.W.\ and Y.Z.\ acknowledge JWST/NIRCam contract to the University of Arizona NAS5-02015. 
This publication is based on work supported by the World Premier International Research Center Initiative (WPI Initiative), MEXT, Japan, KAKENHI (20H00180) through the Japan Society for the Promotion of Science. This work was supported by the joint research program of the Institute for Cosmic Ray Research (ICRR), the University of Tokyo.
J.M.H. is supported by the JWST/NIRCam Science Team contract to the University of Arizona, NAS5-02015, and JWST Program 3215.
K.K. acknowledges support from VILLUM FONDEN (71574). The Cosmic Dawn Center is funded by the Danish National Research Foundation under grant no. 140.



%

\vspace{5mm}





\bibliography{papers}{}
\bibliographystyle{aasjournal}



\end{document}